\documentclass{bmcart}

\usepackage{amsthm,amsmath}
\usepackage[numbers,sort&compress]{natbib}
\usepackage[utf8]{inputenc} 
\usepackage{soul}
\usepackage{float}
\usepackage{placeins}
\usepackage{pdflscape}

\usepackage[para,online,flushleft]{threeparttable}
\usepackage{booktabs}
\usepackage{authblk} 
\usepackage[utf8]{inputenc}
\usepackage[titletoc,title]{appendix}
\usepackage{bm}
\usepackage{multicol} 
\usepackage[numbers,sort&compress]{natbib}

\usepackage{xcolor}
\usepackage{soul}
\usepackage{longtable}
\usepackage{tabularray}
\usepackage{url}
\DeclareUnicodeCharacter{0302}{\^}

\usepackage{graphicx,float}
\usepackage{lmodern}
\usepackage{anyfontsize}

\usepackage{float}
\usepackage[para,online,flushleft]{threeparttable}
\usepackage{booktabs}

\usepackage{subfloat}

\graphicspath{{Images/}}

\startlocaldefs
\endlocaldefs

\begin{document}

\begin{frontmatter}

\begin{fmbox}
\dochead{Simulation study}


\title{Evaluating the effect of different non-informative prior specifications on the Bayesian proportional odds model in randomised controlled trials: a simulation study}


\author[
  addressref={aff1,aff2},                   
  corref={aff1,aff2},                       
  email={chris.selman@unimelb.edu.au}   
]{\inits{C.J}\fnm{Chris J.} \snm{Selman}}
\author[
  addressref={aff1,aff2},
  email={katherine.lee@mcri.edu.au}
]{\inits{K.J}\fnm{Katherine J.} \snm{Lee}}
\author[
  addressref={aff3,aff4}
]{\inits{M.D.}\fnm{Michael} \snm{Dymock}}
\author[
  addressref={aff10}
]{\inits{I.M.}\fnm{Ian C.} \snm{Marschner}}
\author[
  addressref={aff5,aff6}
]{\inits{S.Y.C}\fnm{Steven Y.C.} \snm{Tong}}
\author[
  addressref={aff3,aff9}
]{\inits{M.J.}\fnm{Mark} \snm{Jones}}
\author[
  addressref={aff1,aff7, aff8},
  email={robert.mahar@mcri.edu.au}
]{\inits{R.K}\fnm{Robert K.} \snm{Mahar}}

\address[id=aff1]{
  \orgdiv{Clinical Epidemiology and Biostatistics Unit},             
  \orgname{Murdoch Children's Research Institute},          
  \city{Parkville},  
   \state{VIC},
  \postcode{3052},
  \cny{Australia}                                    
}
\address[id=aff2]{%
  \orgdiv{Department of Paediatrics},
  \orgname{University of Melbourne},
  \city{Parkville},  
   \state{VIC},
  \postcode{3052},
  \cny{Australia}                                    
}
\address[id=aff3]{%
  \orgdiv{Wesfarmers Centre of Vaccines and Infectious Diseases},
  \orgname{The Kids Research Institute Australia},
  \city{Nedlands},  
   \state{WA},
  \postcode{6009},
  \cny{Australia}                                    
}
\address[id=aff4]{%
  \orgdiv{School of Population and Global Health},
  \orgname{The University of Western Australia},
  \city{Nedlands},  
   \state{WA},
  \postcode{6009},
  \cny{Australia}                                    
}
\address[id=aff10]{%
  \orgdiv{NHMRC Clinical Trials Centre},
  \orgname{University of Sydney},
  \city{Camperdown},  
   \state{NSW},
  \postcode{2050},
  \cny{Australia}                                    
}
\address[id=aff5]{%
  \orgdiv{Department of Infectious Diseases},
  \orgname{University of Melbourne, at the Peter Doherty Institute for Infection and Immunity},
  \city{Parkville},  
   \state{VIC},
  \postcode{3052},
  \cny{Australia}                                    
}
\address[id=aff6]{%
  \orgdiv{Victorian Infectious Diseases Service},
  \orgname{The Royal Melbourne Hospital, at the Peter Doherty Institute for Infection and Immunity},
  \city{Parkville},  
   \state{VIC},
  \postcode{3052},
  \cny{Australia}                                    
}

\address[id=aff9]{%
  \orgdiv{Sydney School of Public Health, Faculty of Medicine and Health},
  \orgname{University of Sydney},
  \city{Sydney},  
   \state{NSW},
  \postcode{2006},
  \cny{Australia}                                    
}
\address[id=aff7]{%
  \orgdiv{Centre for Epidemiology and Biostatistics, Melbourne School of Population and Global Health},
  \orgname{University of Melbourne},
  \city{Parkville},  
   \state{VIC},
  \postcode{3052},
  \cny{Australia}                                    
}
\address[id=aff8]{%
  \orgdiv{Methods and Implementation Support for Clinical and Health Research Hub},
  \orgname{University of Melbourne},
  \city{Parkville},  
   \state{VIC},
  \postcode{3052},
  \cny{Australia}                                    
}



\end{fmbox}


\begin{abstractbox}

\begin{abstract} 
\small	
\parttitle{Background} 
Ordinal outcomes combine multiple distinct ordered patient states into a single endpoint and are commonly analysed using proportional odds (PO) models in clinical trials. When using a Bayesian approach, it is not obvious what the influence of a `non-informative' prior is in the analysis of a fixed design or on early stopping decisions in adaptive designs.

\parttitle{Methods} 
This study compares different non-informative prior specifications for the Bayesian PO model in the context of both a two-arm trial with a fixed design and an adaptive design with an early stopping rule. We conducted an extensive simulation study, varying the effect size, sample size, number of categories and distribution of the control arm probabilities. The effect of the prior specifications is also compared using data from the Australian COVID-19 Trial.

\parttitle{Results} 
Our findings indicate that the choice of prior specification can introduce bias in the estimation of the treatment effect, particularly when control arm probabilities are right-skewed. The R-square prior specification resulted in the smallest bias and increased the likelihood of appropriately stopping early when there was a treatment effect. However, this specification exhibited larger biases when control arm probabilities were U-shaped when there was an early stopping rule. Dirichlet priors with concentration parameters close to zero resulted in the smallest bias when probabilities were right-skewed in the control arm, but were more likely to inappropriately stop early for superiority when there was no treatment effect and an early stopping rule. 

\parttitle{Conclusions} 
The specification of non-informative priors in Bayesian adaptive trials with ordinal outcomes has implications for treatment effect estimation and early stopping decisions. We recommend the careful selection of priors that consider the possible distribution of control arm probabilities and that sensitivity analyses to the prior be conducted.

\end{abstract}




\end{abstractbox}
%

\end{frontmatter}



\section*{Introduction}
Randomised controlled trials (RCTs) aim to answer causal questions by comparing outcomes from participants randomised between a control treatment and at least one intervention. One type of outcome that is becoming increasingly popular in trials is an ordinal outcome \cite{selman2024statistical}. Ordinal outcomes combine multiple distinct ordered patient states into a single endpoint. The categories that make up the ordinal outcome must be monotonically ordered but there may not necessarily be an equal `distance' between proximate categories. Although ordinal outcomes may be less familiar to researchers than other outcome types, they are commonly used in many medical settings. For example, the modified-Rankin scale \cite{bonita1988recovery} is typically used to measure outcomes in patients who suffer from stroke or traumatic brain injury with categories ranging from 0 (no symptoms) and 6 (dead), with increasing severity of disability in between. Ordinal outcomes became particularly prominent during the initial stages of the COVID-19 pandemic, when the World Health Organization developed the Clinical Progression Scale to measure distinguished health states for patients with COVID-19 \cite{mathioudakis2020outcomes} that was anchored at asymptomatic and uninfected, and death. Despite being relatively common, there is still a lack of understanding of the interpretation and appropriate analyses of ordinal outcomes in RCTs \cite{selman2024statistical}.  

There are many approaches available to analyse ordinal outcomes in RCTs. The most common is a proportional odds (PO) model, which estimates a proportional (or common) odds ratio (OR) across each binary split (or `cut-point') of the ordinal scale \cite{walker1967estimation, mccullagh1980regression, harrell2015regression, selman2024statistical}. The proportional OR represents the size of the shift in the distribution of the ordinal scale towards a favourable (or unfavourable) outcome for an intervention relative to the comparator, which can be used to summarise the efficacy of an intervention. 

Alongside the increased use of ordinal outcomes, the COVID-19 pandemic also necessitated efficient trial designs that could expedite treatment decisions \cite{stallard2020efficient, perkins2020recovery, florescu2022effect}. This included the use of adaptive designs that allow prespecified adaptations or changes to the design based on the accruing data from the trial. Some examples of adaptations that can be incorporated include early stopping, sample size re-estimation, response adaptive randomisation, subgroup selection, or some combination thereof \cite{kairalla2012adaptive}.  

Early stopping rules, particularly for superiority, are commonly used in trials that use ordinal outcomes \cite{selman2024statistical}. Early stopping for superiority allows a trial to be stopped following a pre-specified interim analysis, if there is sufficient evidence that an intervention is superior to the comparator (or sometimes, other interventions), avoiding future patients being allocated to inferior treatments. A trial with such stopping rules can reduce the anticipated duration and sample size compared to a trial with a fixed design \cite{mcmurray2014angiotensin}. 

Bayesian methods are commonly used for analysing adaptive trials that allow for early stopping since they are naturally suited to multiple looks at the data, complicated models can be fit fairly easily, and prior distributions provide a principled way of updating inferences at each interim analysis \cite{berry2006bayesian, bast2010holland, berry2005introduction}. In Bayesian inference, the data are combined with prior distributions or beliefs about the parameters of interest to obtain a posterior distribution for the parameters \cite{gelman2013philosophy}. The choice of prior distribution has a direct influence on the posterior distribution, particularly if an informative prior is chosen or the sample size is small, which can in turn affect the final estimate of the treatment effect. In an adaptive trial, the decision to stop a trial early for superiority is typically guided by the posterior probability that the intervention is superior to one or more comparators which can again be affected by the choice of prior. For example, if an informative prior strongly favours the intervention over the comparator, early stopping decisions may be (possibly inappropriately) accelerated which would increase the risk of stopping for success even if the intervention is not superior.

Non-informative priors are often used in the analysis of trials to reflect clinical equipoise and to allow the inference to be guided by the data. There are, however, numerous non-informative priors that can be used for the treatment effect and any other parameters in the model, and it is often unclear how `non-informative' these priors may be for a complex analysis model, particularly in the analysis of ordinal outcomes. For example, although a diffuse prior centered around no treatment effect, such as a $\text{Normal}(0, \sigma)$ prior distribution with large $\sigma$, may be a natural choice for the treatment effect parameter, the influence on the posterior distribution is not well understood in the context of a PO model. For large $\sigma$, the inverse-logit transformation of the prior distribution pushes the mass towards the lower and upper bounds, creating a `bathtub' like distribution. Instead, specifying a smaller $\sigma$ may be more appropriate. Furthermore, PO models for ordinal outcomes use ordered cut-point parameters to distinguish between consecutive (outcome) categories that represent cumulative log-odds for the control group which also require prior distributions. There has been some methodological research surrounding the choice of priors however, this research has primarily focused on the choice for the prior when the outcome is continuous. In the context of PO models, it has been suggested that for the model cut-points a prior should be chosen that preserves the parameter ordering, such as a Dirichlet distribution \cite{mckinley2015bayesian, james2021bayesian} or Normal distribution. However the impact that different non-informative priors have on the estimation of the treatment effect has not been explored in trial settings. Given the increased use of ordinal outcomes in trials that use Bayesian PO models and the numerous possible choices of prior distributions available, it is imperative to understand the impact that non-informative priors have on the posterior distribution for the treatment effect and whether they are truly `non-informative', particularly in trials with interim analyses where the sample size may be small.

In this paper, we aim to evaluate the effect of different non-informative prior specifications on the estimation of the treatment effect from a Bayesian PO model. We focus on the context of a two-arm trial with 1:1 ratio of allocation that uses an ordinal outcome under 1) a fixed design that has a single analysis at the end of the study and, 2) an adaptive design with an early stopping rule for superiority at a single interim analysis. This is investigated using a simulation study and a case study of the Australasian COVID-19 Trial (ASCOT) trial data. As part of the simulation study we assess the impact of the following factors on the estimation of the treatment effect and posterior probability of intervention superiority where the target estimand is a proportional OR:

\begin{enumerate}
    \item Magnitude and direction of the proportional OR.
    \item Trial sample size.
    \item Number of categories of the ordinal outcome.
    \item Distribution of the control arm outcome probabilities. 
\end{enumerate}

This paper proceeds as follows: we first define key notation, provide an overview of the PO model and the estimand of interest, before outlining candidate non-informative prior distributions. Next we describe and present the results of the simulation study followed by the results of the ASCOT case study \cite{mcquilten2023anticoagulation}. We conclude by summarising the key results and provide guidance for specifying priors for Bayesian PO models.

\section*{Methods}
\subsection*{Notation}

The target estimand is the proportional OR that represents the relative increase in odds of a more favourable outcome for an intervention compared to a control arm. We start by outlining our notation:

\begin{itemize}
    \item $J$ is the number of categories in the ordinal outcome;
    \item $\beta$ is the proportional log-OR;
    \item $b \in \{c,t\}$ represents the control arm $c$ and intervention arm $t$;
    \item $\boldsymbol{\pi}_b$ is a vector of ordinal outcome category probabilities for participants receiving arm $b$. That is:
\begin{equation}
\boldsymbol{\pi_b}=\left(\pi_{b 1}, \pi_{b 2}, \ldots, \pi_{b J}\right)^{T}
\end{equation}
\item $\boldsymbol{\alpha} = (\alpha_{2},...,\alpha_{J})$ is a vector of model intercepts representing the cumulative log-odds for the control group;
\item $a = 1, 2$ indicate the interim and final analysis times;
\item $\boldsymbol Y_{ab}$ is the vector of observed ordinal outcome for participants $n = 1, 2, ... , n_{a}$ allocated to arm $b$ at analysis $a$.
\item $x_{i} \in \{0, 1\}$ is an indicator denoting assignment to intervention, where 0 = control and 1 = intervention. 

\end{itemize}

\subsection*{Proportional odds model}
The PO model can be used to estimate a treatment effect for an ordinal outcome with more than two categories. The model estimates a proportional log-OR ($\beta$) that is assumed to be constant across all binary splits of the ordinal scale, where for $j = 2, 3, ..., J$:
\begin{equation} \label{eq:cumulative_logit}
\log \frac{P(Y_{ab} \geq j)}{P(Y_{ab} < j)}=\mathrm{logit}\big[P(Y_{ab} \geq j)] =\alpha_{j}+ \beta \mathbf{x}
\end{equation}

When using this model for Bayesian inference, a prior must be specified for the baseline cumulative log-odds $\boldsymbol{\alpha}$ and for the proportional log-OR $\beta$. We outline some options of possible priors for these parameters in a PO model in the following section.

\subsection*{Specification of non-informative priors for $\beta$}

A non-informative prior on the treatment effect parameter $\beta$ is usually specified to be symmetric around no effect (i.e., zero) with sufficient variability to encapsulate the appropriate range of potential effects if a non-informative prior is desired \cite{gelman1995bayesian, gelman2008weakly, sarma2020prior}. One common choice is a Normal distribution with a large variance parameter \cite{selman2024statistical}. Although this prior appears to be a reasonable choice on the log-odds scale, when transforming the distribution using the inverse-logit transformation, the mass is pushed towards the bounds at zero and one, creating a `bathtub' like distribution. Instead, using a smaller variance parameter, although appearing more informative, may still be rather non-informative. Other symmetric and potentially non-informative prior distributions that could be specified for the treatment effect include the Student-$t$ distribution \cite{sarma2020prior} (which approaches the shape of a Normal distribution when the degrees of freedom becomes large but has fatter tails with smaller degrees of freedom to reflect a wider range of prior beliefs), a Cauchy distribution (a $t$-distribution with one degree of freedom), or a Laplace distribution. The Laplace distribution is also known as the double-exponential distribution and is symmetric with a sharp peak at its centre and can be motivated as a scale mixture of Normal distributions. This distribution has fairly flat tails to mimic a non-informative prior when an appropriate value is specified for the standard deviation \cite{rochfordprior}.

An alternative prior specification for the treatment effect parameter $\beta$ is to put a prior on the beliefs about the location of the $R^2$ \cite{rsquare1} (the proportion of variance in the ordinal outcome that is explained by the intervention) assuming a continuous latent variable specification for the ordinal outcome. This involves putting a prior on the location of $R^{2}$, which can be achieved using a Beta distribution where the first shape hyperparameter is set to equal to half the number of predictors in the model \cite{rsquare1} and the second shape hyperparameter is greater than zero. When both shape hyperparameters are equal (in this case, also 0.5), the prior mean and median of the $R^{2}$ is 0.5 and the distribution is symmetric. Like the normal prior with a large variance, this specification pushes most of the probability density toward zero and one. Choosing a value of the second shape hyperparameter that is not close to zero or one would be a reasonable diffuse prior since the prior belief does not hinge on whether the intervention has a strong effect on the outcome (if close to one), and whether there is high certainty that the intervention has no effect on the outcome (if close to zero). It can be unclear how to specify a prior value for the $R^2$ in a PO model as $R^2$ refers to the proportion of variance in the underlying continuous latent variable that is attributable to the predictors under a linear model, and typically, the $R^2$ is lower in a PO model compared to a linear model with a continuous outcome. However, the smaller the $R^2$, the smaller the prior correlations among the ordinal outcome and intervention variable(s) are, and the more concentrated near zero the prior distribution is for the regression coefficient (i.e. the proportional log-OR).

In the current study, we compare the estimation performance of the PO model using the following priors on $\beta$: 

\begin{enumerate}
    \item $\beta \sim N(\mu = 0, \sigma^2 = 100^2)$
    \item $\beta \sim N(\mu = 0, \sigma^2 = 2.5^2)$
    \item $\beta \sim t(df = 1)$ (equivalent to a Cauchy distribution)
    \item $\beta \sim$ Laplace$(\mu = 0)$, with SD = 100
    \item $\beta \sim$ Laplace$(\mu = 0)$, with SD = 2.5
    \item Mean $R^2$ = 0.5, equivalent to a Beta(0.5, 0.5) distribution
\end{enumerate}

\subsection*{Specification of non-informative priors for $\boldsymbol{\alpha}$}

Although we would expect the prior used for the treatment effect to have the largest effect on the posterior distribution for the treatment effect, the prior used for the cut-points may also be important. The Dirichlet distribution is a natural choice for the (implicit) prior on the model cut-points because the concentration parameters can be interpreted as prior counts, i.e. prior probabilities for each of the $J$ categories at the covariate means \cite{gelman1995bayesian}. For instance, if $\alpha_{j} = 1$ for all $ j \in \{1,2,\ldots,J\}$, the Dirichlet prior is uniformly distributed across the categories. This implies a prior belief of one observation in each of the $J$ ordinal categories when the predictors are at their sample means, indicating very weak prior knowledge that no category has probability zero. The Dirichlet prior distribution may be appropriate when modeling ordinal outcomes given its multivariate generalisation of the Beta distribution and that it is the conjugate prior of the categorical and multinomial distribution \cite{tu2014dirichlet}. An alternative choice is Jeffrey's fixed reference prior where $\alpha_{j} = 0.5$ for all $j \in \{1,2,\ldots,J\}$ \cite{james2021bayesian}. This prior maximises the `distance' (based on the Kullback–Leibler divergence, a measure of how one probability distribution is different to another probability distribution) between the prior and the posterior, implying that the data has the largest possible effect on the posterior (and that the prior has the least possible effect on the posterior). An alternative is to specify a normal distribution for each cut-point that preserves the ordering in the model. 

In this paper we investigate the following priors for the intercept terms: 

\begin{enumerate}
    \item Uniform Dirichlet: $\boldsymbol{\alpha} \sim Dir(\boldsymbol{1})$ (sets of baseline risk are equally likely for each category).
    \item Multivariate Jeffrey's reference prior for the cut-points: $\boldsymbol{\alpha} \sim Dir(\boldsymbol{0.5})$.
    \item $\boldsymbol{\alpha} \sim Dir(\boldsymbol{0.001})$.
    \item $\boldsymbol{\alpha} \sim Dir(\frac{1}{J})$ (objective reciprocal prior specification)
    \item $\alpha_j$ $\sim N(0, 100^2)$
\end{enumerate}

\section*{Simulation study}
We conducted a simulation study to evaluate the impact that different non-informative priors for $\boldsymbol{\alpha}$  and $\beta$ have on the estimation of the treatment effect under a range of distributional, sample size, trial design, and effect size scenarios. In this section, we outline the data generating mechanisms, analysis methods, and performance measures used in the simulation study. 

\subsection*{Data-generating mechanisms}
The data generation process assumes a true underlying proportional OR in each trial. We set up a hypothetical scenario where we emulate a two-arm trial with a fixed design, and another with a single interim analysis once half of the outcome data is available with the potential to stop early if the intervention is declared superior. Intervention assignment was generated using $X_{i} \sim \mathrm{Bern}(0.5)$ - representing simple randomisation with an equal allocation ratio. The ordinal outcome data was then simulated by taking a random sample from a multinomial distribution for the control and intervention group respectively given the respective probabilities as described below. We restrict our focus to ordinal outcomes with $J \in \{4, 10, 30\}$ categories to reflect the wide range of ordinal outcomes that are used in practice. 

We varied the distribution of the probabilities in each outcome category in the control group to reflect real-world scenarios. First, we set the probabilities in each outcome category to follow a skewed relationship as observed in the distribution of two ordinal outcomes that were both days alive and free of hospitalisation/ventilation 28-days post randomisation respectively in the ASCOT trial. Here, the probabilities were primarily concentrated to categories at the beginning of the outcome scale, exhibiting positive skewness across the distribution of the categories. Second, we defined probabilities in each outcome category to follow a U-shape distribution which is commonly observed in stroke trials that use the modified-Rankin scale where patients tends to fall in either the first two or three categories and the last two categories (depending upon the population), as was observed in the ESCAPE-NA1 trial \cite{hill2020efficacy}. We set probabilities in each category of the ordinal outcome in the control group by partitioning the sample space of a $\theta \sim$ Beta$(\alpha, \beta)$ distribution into $J$ equal partitions where the control probabilities for each category were set to equal the cumulative probability in each partition. Finally, we considered a scenario where the probabilities in each outcome category were equal.

Next we generated the cumulative log-odds for the intervention group assuming proportional odds, i.e. $\mathrm{logit}(P(Y\geq j)) = \alpha_{j} + \beta$ where we explored an effect size of log(OR) = $\mathrm{log}(1)$ (no effect), $\mathrm{log}(1.10)$ (small effect) and $\mathrm{log}(1.50)$ (moderate effect size) to reflect similar proportional ORs to that observed in the ASCOT trial. The intervention outcome probabilities $\boldsymbol \pi_t$ were then calculated using the inverse logit function. These probabilities along with the control group probabilities were used to generate the ordinal outcome.

Data was simulated for a total sample size of $n_{obs} \in \{100,500\}$ records for each scenario under consideration. For simplicity, we assumed that the outcome is observed immediately. An interim analysis occurred at an information fraction of one-half, i.e. at a sample size of 50 or 250. The stopping rule was guided by the Bayesian posterior probability that the intervention was superior to the control arm. Specifically, superiority of an intervention was assessed using the posterior probability that the proportional log-OR is greater than zero exceeds 0.99 (i.e. $P(\beta > 0) > 0.99$) at the interim analysis to align with more stringent stopping rules at an interim, then superiority was triggered and the trial was stopped early at the interim analysis. If the trial continued to the maximum sample size, superiority was declared if $P(\beta > 0) > 0.95$. For the fixed design, superiority was declared if $P(\beta > 0) > 0.95$. We note that the two designs are not comparable due to different Type I error rates, although the objective of this study is not to compare the designs, rather the impact of using different priors on the treatment effect given the study design. In total, 108 scenarios were considered in this simulation study. 

\subsection*{Analysis methods} 
We estimated the parameters of the PO model (Equation 2) using Hamiltonian Monte Carlo (HMC) sampling, a Markov Chain Monte Carlo (MCMC)-sampling-based algorithm, to obtain the posterior distribution for the treatment effect using the different prior specifications for $\boldsymbol{\alpha}$  and $\beta$ outlined above. We  initially assess the impact of changing the prior for the treatment effect $\beta$ for a fixed prior for $\boldsymbol{\alpha}$ ($\boldsymbol{\alpha} \sim Dir(\boldsymbol{1})$), and then assess the impact of changing the prior for the cut-points but holding the prior for $\beta$ fixed $\beta \sim N(0, 100^2)$. The median of the posterior distribution for $\beta$ was used to determine performance metrics. The HMC method was implemented in the R programming environment \cite{rstudio} and Stan \cite{carpenter2017stan}, the latter of which is a probabilistic programming language that specifies statistical models. The R packages \textit{posterior} \cite{posteriorpackage} and \textit{rstanarm} \cite{Goodrich2020} were used to summarise posterior distributions.

An iterative process was used to determine the number of simulations, $n_{sim}$ for each scenario, to ensure that the required maximum tolerable upper bound of the Monte Carlo standard errors (MCSEs) for each performance measure was less than 0.05 \cite{kelter2023bayesian}. We calculated the jackknife-after-bootstrap MCSE for each parameter using the method described in Koehler et al \cite{koehler2009assessment}. All scenarios achieved the upper bound when $n_{sim} = 1000$.

A burn-in of 5,000 iterations and a chain length of 10,000 samples per chain were used across four chains, with no thinning, resulting in a total of 20,000 post-warmup samples from the posterior distribution for each scenario. To evaluate the reliability of the posterior distributions, convergence diagnostics were recorded, including the effective sample size (ESS), the potential-scale reduction factor ($\hat{R}$), and the number of divergent transitions, which signal difficulties in properly exploring the parameter space. Divergent transitions, which occur when the HMC sampler encounters extreme curvature in the target distribution, were minimized or eliminated by adjusting the average target acceptance probability rate and increasing the maximum tree depth. If divergent transitions were present, a sensitivity analysis that removed the analysis method where such transitions occurred was performed to assess the robustness of the conclusions. 

\subsection*{Performance measures}

The performance measures of interest were the overall bias, relative bias, coverage and mean-squared error of the proportional log-OR, relative to the value used in the data generation \cite{morris2019using}. Relative bias was calculated on the OR scale. For each scenario, we also estimated the posterior probability that the intervention was superior (whether this be at the interim or maximum sample size), the proportion of trials that declared superiority, and among the trials that incorporated an early stopping rule, the proportion of trials that stopped early.

\section*{Simulation study results}
In this section, we summarise the results from the simulation study. There was a similar pattern of results for the fixed and adaptive designs so we focus on the latter and present the results for the fixed design in the supplementary material, unless where patterns were different.

\subsection*{Specification of priors for $\beta$}
There was substantial bias in the estimated treatment effect when the outcome distribution in the control arm was right-skewed for all prior distributions apart from a prior specification on the R-squared, with the relative bias increasing for increasing number of categories (see Figure 1 and Figure S2 in Supplementary Material 1). However, the bias in the treatment effect was generally minimal irrespective of the prior specification when the control probabilities were U-shaped or uniform for the fixed design. The exception was when there was moderate effect size and sample size was small for the Laplace (small SD), Cauchy and R-square prior specifications when there was a  positive bias, particularly for the adaptive designs. Coverage showed a similar pattern being close to nominal coverage level ($95\%$) when the control probabilities were U-shaped or uniform, but with under-coverage when they were skewed and there were 30 categories. This was the case for both fixed and the adaptive design. Looking at the average posterior probability that the intervention was superior (see Figures S9 and S10 in Supplementary Material 1), it appears that with smaller effect size, the posterior probability decreased with higher number of categories in the ordinal outcome for all priors. In particular, when there were skewed probabilities in the control arm and a small sample size, it appears that the posterior probability was slightly higher when a prior on the R-squared was used compared with the alternative priors, which was the case for both the fixed and adaptive designs. When the control arm probabilities were U-shaped, the proportion of trials that declared superiority did not appear to differ substantially with the choice of prior, although there was a slightly higher proportion of trials declaring superiority with the Normal, Laplace and Cauchy prior compared to when a prior specified using the R-squared was used. Similar results were observed for the fixed design. 

A higher proportion of trials with an adaptive design declared superiority when the R-squared prior was used with skewed control arm probabilities when there was a small and moderate effect size, particularly with increasing number of categories in the ordinal outcome (Figure 2c). When there were skewed control arm probabilities, the R-squared prior results in the highest proportion of trials that would have stopped early, but when the control arm probabilities were U-shaped the R-squared prior resulted in the fewest trials stopping early. In the latter context, specifying Normal or Laplace priors with large standard deviations had the highest proportion of trials that stopped early particularly with the larger effect sizes. It appears that trials are less likely to stop early when the ordinal outcome is right-skewed in the control arm and has a larger number of categories, regardless of the prior specified for the treatment effect, with the exception of when the prior is specified on the R-squared. 

Of note, the posterior distributions for all target parameters of interest across all scenarios appeared to converge to their stationary distribution, with $\hat{R} < 1.01$ for all priors and all scenarios. The bulk and tail effective sample sizes were all above the recommendation of 100 per chain \cite{goodrich2020rstanarm,vehtari2021rank}, indicating good mixing and low autocorrelation. The MCSE of all performance measures were less than $0.05$ and are reported in the supplementary material for each prior specification and each scenario (Supplementary Material 2). There were no analyses where there were divergent transitions.

\begin{landscape}
\begin{subfigures}

\begin{figure}[!htbp]
\centering
\caption{Relative bias in the odds ratio for the various scenarios with an adaptive design when varying the prior for the treatment effect }
\small\textsuperscript{OR = odds ratio, SD = standard deviation}

\advance\leftskip-1cm
\advance\rightskip-1cm
\includegraphics[width=1.5\textwidth]{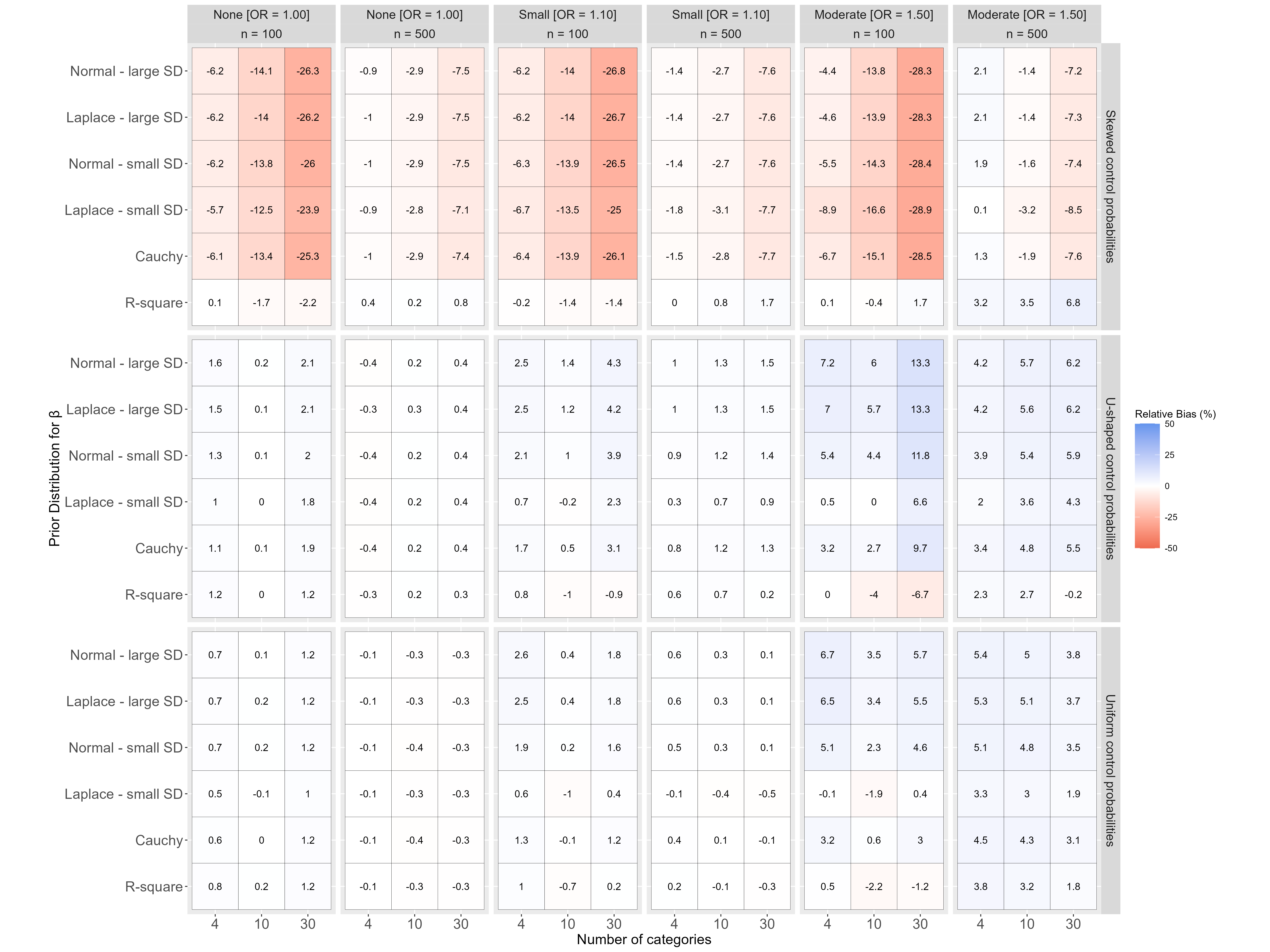}
\end{figure}

\begin{figure}[!htbp]
\centering
\caption{Coverage of the 95\% credible interval for the log-odds ratio for the various scenarios with an adaptive design when varying the prior for the treatment effect}
\small\textsuperscript{OR = odds ratio, SD = standard deviation}

\advance\leftskip-3cm
\advance\rightskip-3cm
\includegraphics[width=1.5\textwidth]{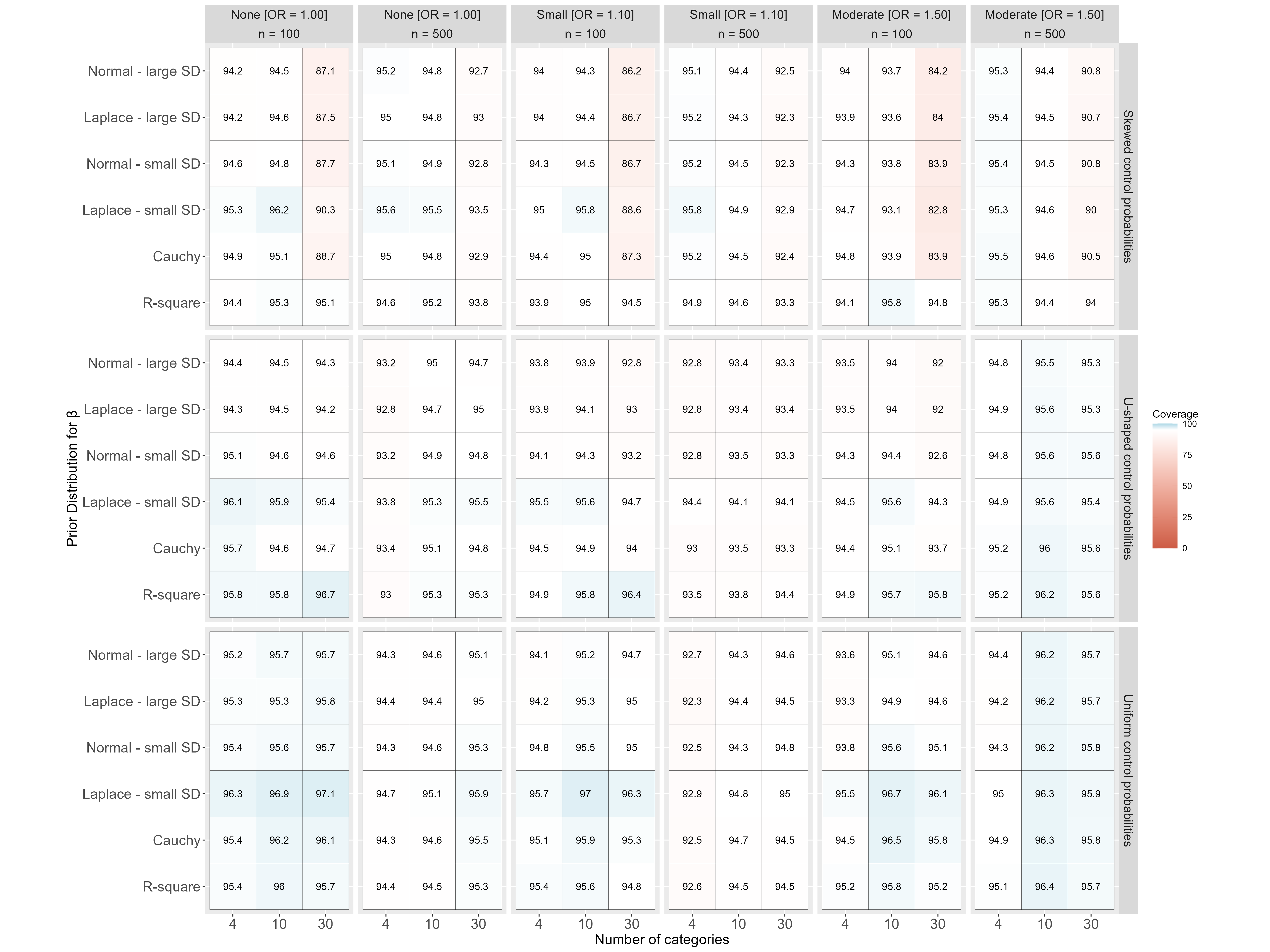}
\end{figure}

\begin{figure}[!htbp]
\centering
\caption{Proportion of trials declaring superiority in an adaptive design when varying the prior for the treatment effect}
\small\textsuperscript{OR = odds ratio, SD = standard deviation}

\advance\leftskip-3cm
\advance\rightskip-3cm
\includegraphics[width=1.5\textwidth]{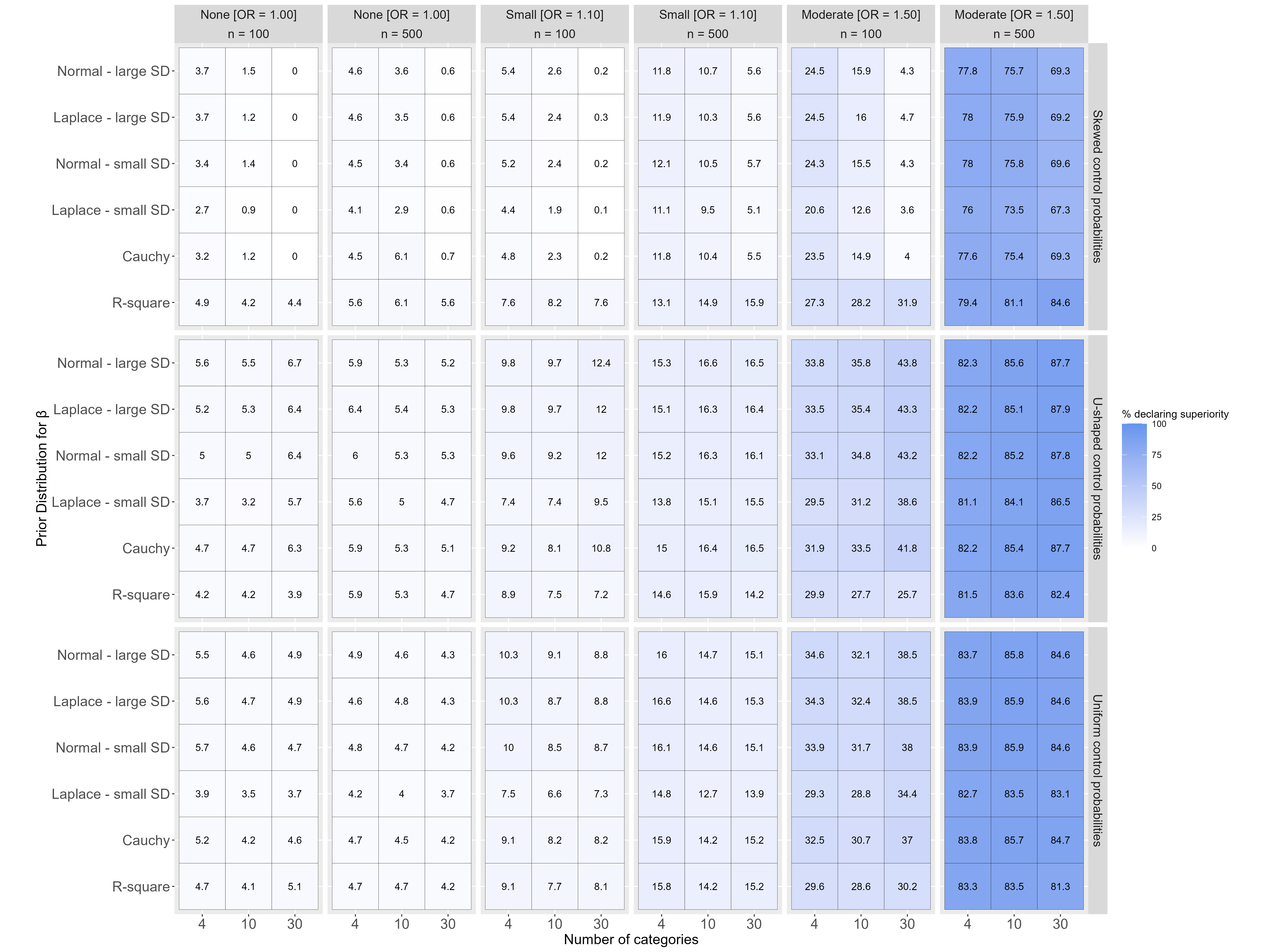}
\end{figure}

\end{subfigures}
\end{landscape}

\subsection*{Specification of implicit priors for $\boldsymbol \alpha$}
Keeping the prior for the treatment effect to be fixed at $\beta \sim N(0, 100)$, as the value of the concentration parameters of the Dirichlet distribution decreased, the relative bias approached zero when the control arm probabilities were right-skewed, regardless of the effect or sample size, for a fixed design (Supplementary Material - Figure S16). In contrast, there was minimal relative bias when the control arm probabilities were right-skewed when a Normal distribution was specified for the cut-points, except when there was a small sample size and the ordinal outcome had 30 categories where there was an overestimation of the treatment effect (relative bias $9\%$ and $11\%$ for skewed and U-shaped probabilities respectively). When the control arm probabilities were U-shaped or uniform, there was minimal bias in the treatment effect for all of the prior specifications except when there were 30 categories and the sample size was small. 

When the design was adaptive, there was a tendency to overestimate the treatment effect with all priors for all scenarios (Figure 3a). This relative bias increased with the number of categories and effect size. The results were similar irrespective of the choice of concentration parameter with larger sample sizes, however, for smaller sample sizes, the priors with a concentration parameter closer to zero resulted in smaller bias for the treatment effect. When the control arm probabilities were right skewed, specifying a Dirichlet using $\boldsymbol{\alpha = 0.5}$ or $\boldsymbol{\alpha = \frac{1}{J}}$ resulted in lower bias that the other priors regardless of the number of categories in the outcome. Coverage was close to the nominal level of $95\%$ for the majority of scenarios (Figure 3b), with the exception of Dirichlet priors where the concentration parameters were closest to zero which resulted in under-coverage. 

When the control arm probabilities were skewed, adaptive designs were less likely to declare superiority and therefore less likely to stop early (Figure 3c) when a Dirichlet prior with $\boldsymbol{\alpha = 1}$ or $\boldsymbol{\alpha = 0.5}$ was used compared to other priors. Importantly, trials were more likely to stop early even when there was no treatment effect when using the Dirichlet prior with a small concentration parameter, with a probability of up to a $18\%$ chance of stopping early in some scenarios.

\begin{landscape}
\begin{subfigures}

\begin{figure}[!htbp]
\centering
\caption{Relative bias in the odds ratio for various scenarios with an adaptive design when varying the implicit prior on the cut-points}
\small\textsuperscript{OR = odds ratio, SD = standard deviation}

\advance\leftskip-1cm
\advance\rightskip-1cm
\includegraphics[width=1.5\textwidth]{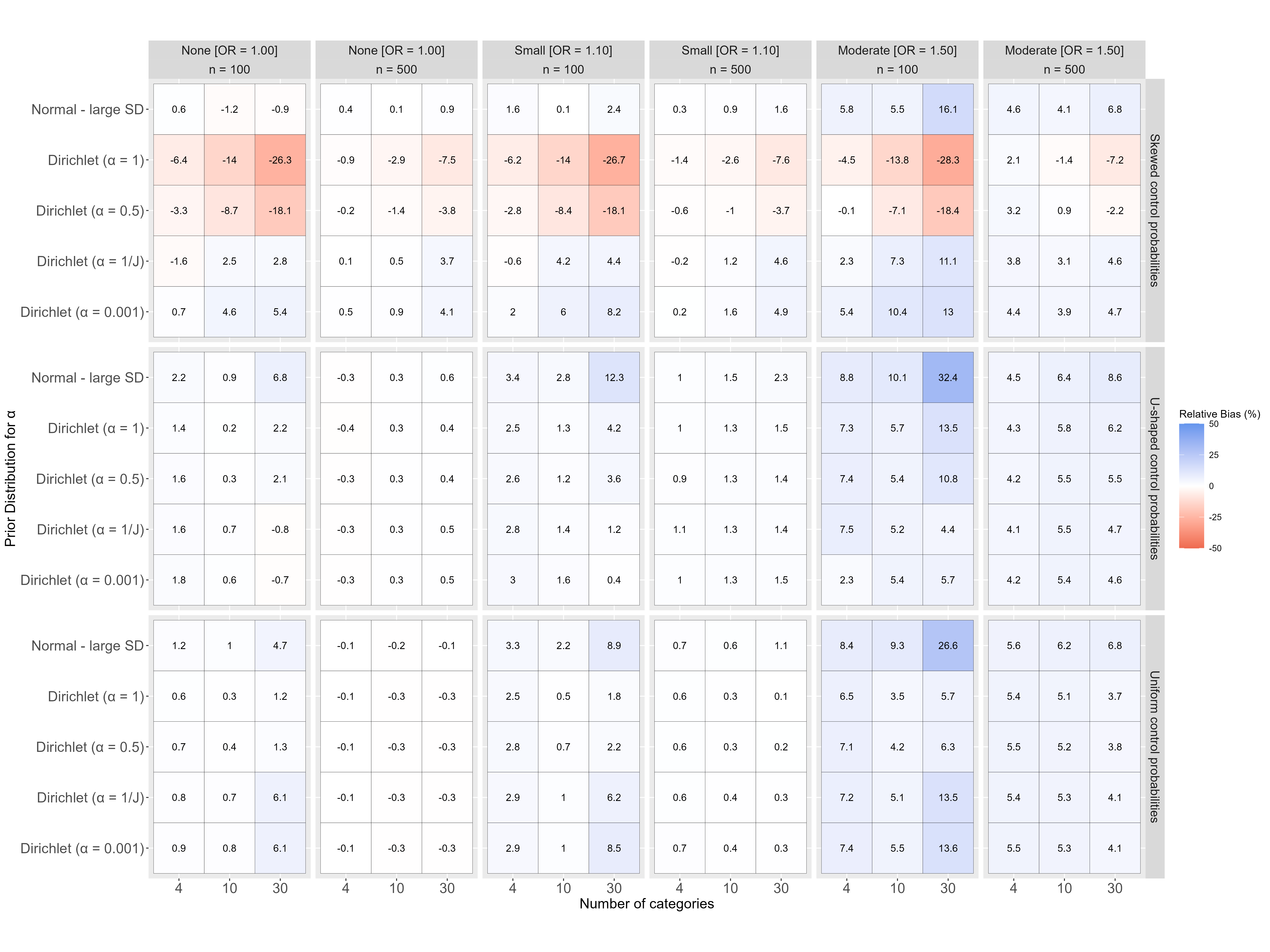}
\end{figure}

\begin{figure}[!htbp]
\centering
\caption{Coverage in the log-odds ratio for various scenarios with an adaptive design when varying the implicit prior on the cut-points}
\small\textsuperscript{OR = odds ratio, SD = standard deviation}

\advance\leftskip-3cm
\advance\rightskip-3cm
\includegraphics[width=1.5\textwidth]{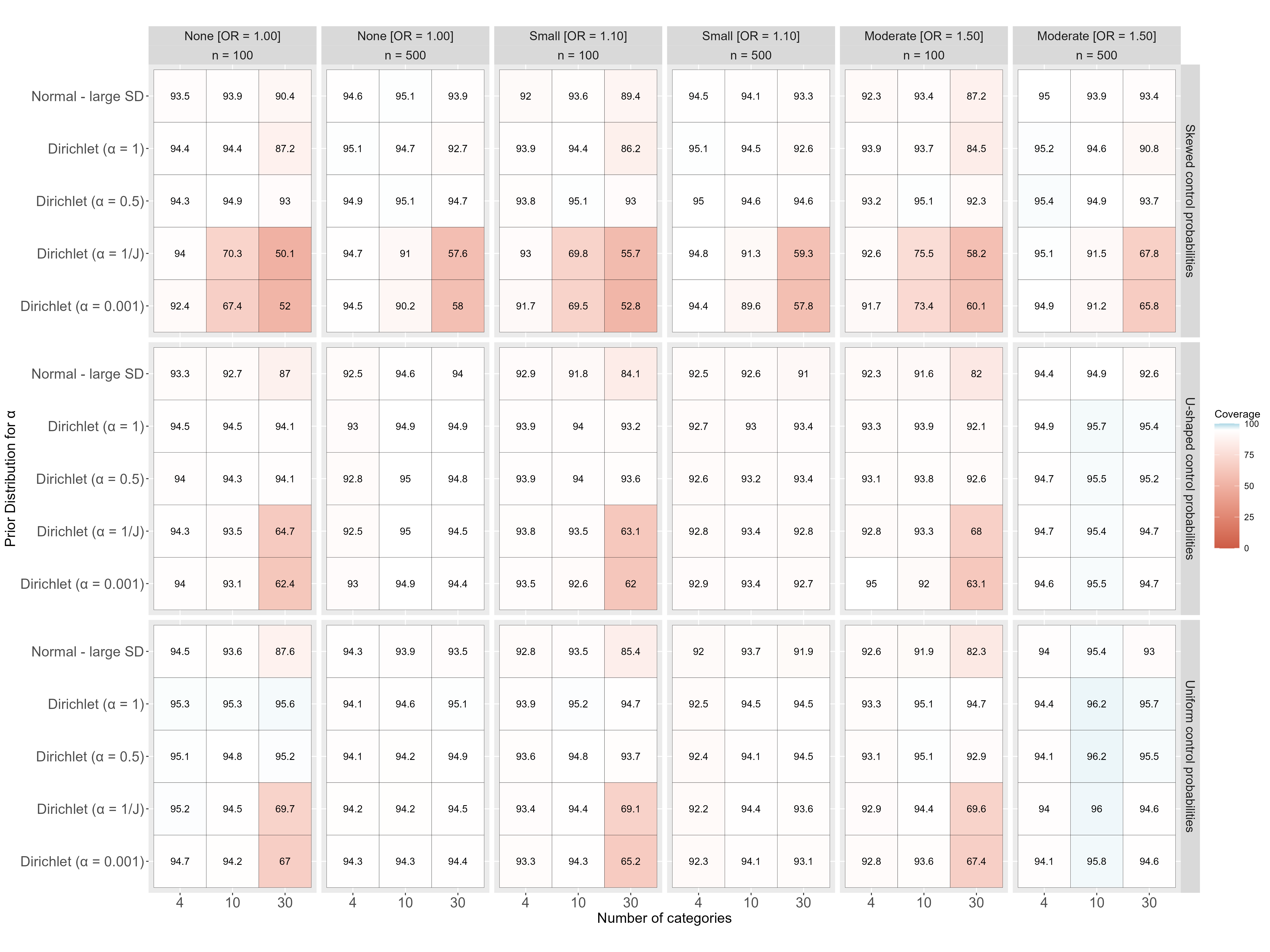}
\end{figure}

\begin{figure}[!htbp]
\centering
\caption{Proportion of trials declaring superiority in an adaptive design when varying the implicit prior on the cut-points}
\small\textsuperscript{OR = odds ratio, SD = standard deviation}

\advance\leftskip-3cm
\advance\rightskip-3cm
\includegraphics[width=1.5\textwidth]{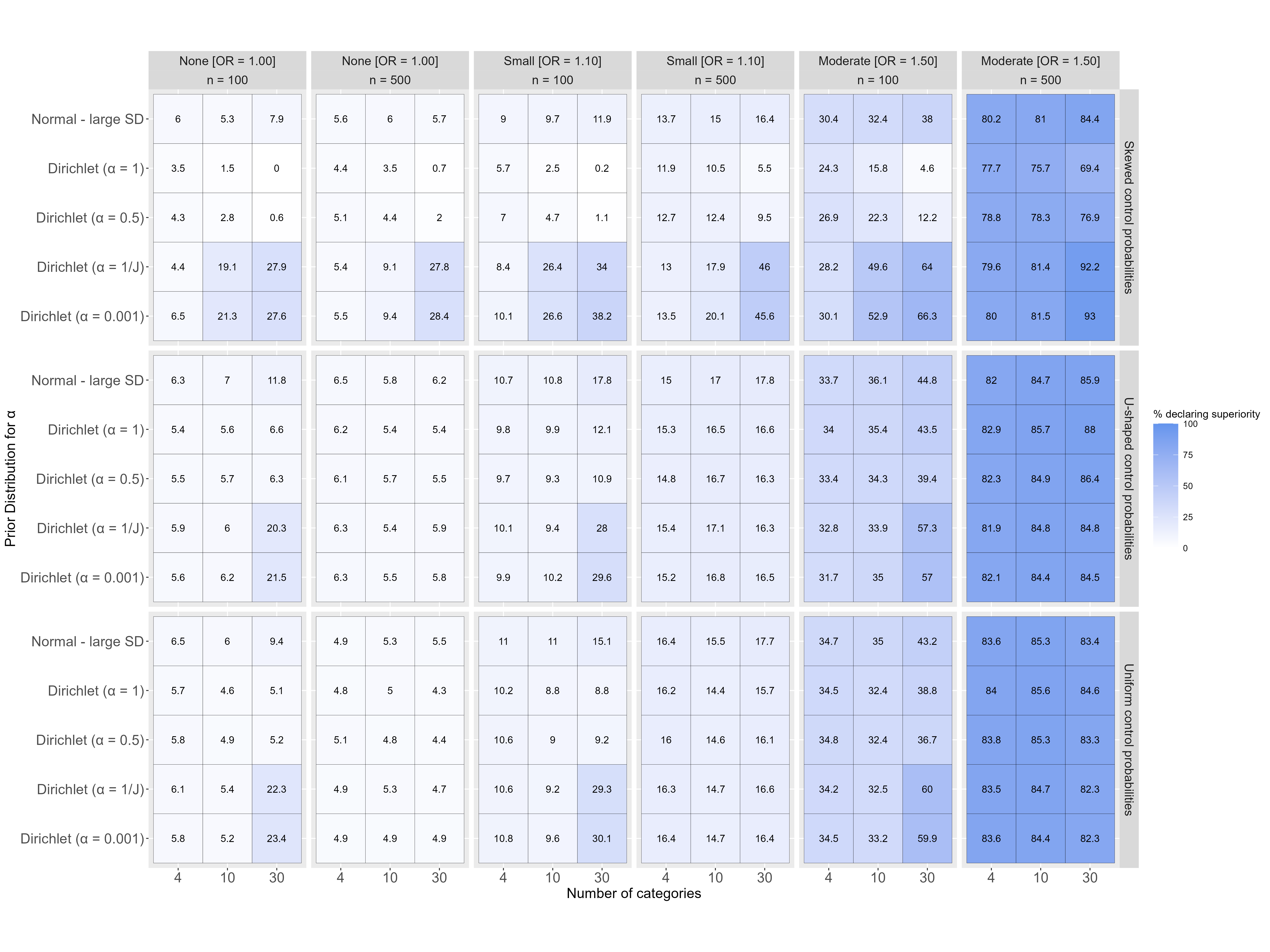}
\end{figure}

\end{subfigures}
\end{landscape}

When altering the prior on the cut-points, divergent transitions sometimes occurred, despite increasing the target acceptance probability rate and maximum tree-depth to reduce the number that occurred. Divergent transitions were particularly common when the Dirichlet prior was used with small concentration parameters when the sample size was small and the outcome had with 30 categories. We present the results with and without divergent transitions given some results had high $\hat{R}$ and very small effective sample sizes (see Supplementary Material 3). When the simulated datasets resulting in divergent transitions were removed, all estimates of the $\hat{R} $ were $< 1.01$, effective sample sizes were sufficiently large, and the MCSE for the performance measures were less than $0.05$. 

\section*{The ASCOT case study}
ASCOT was an adaptive platform trial that evaluated the efficacy of multiple interventions for hospitalised, non-critically ill COVID-19 patients. It evaluated interventions across three intervention domains: anticoagulation, antiviral, and therapeutic antibody domains, each comparing a different set of interventions to standard of care within that domain. The primary outcome was death or the need for organ support by day 28 (a binary outcome), but the trial had four secondary ordinal endpoints (described below). We focus on the anticoagulation domain, which trialed low-dose thromboprophylaxis with low-molecular-weight heparin (LMWH), intermediate-dose LMWH, and a low-dose LMWH with once-daily aspirin. The aspirin arm was discontinued during the trial due to external evidence, and a new arm with therapeutic-dose anticoagulation was introduced. The trial ended early because of funding limitations, slowing participant recruitment, and a recommendation from the data safety and monitoring committee to stop therapeutic anticoagulation due to lack of effectiveness.

ASCOT had four ordinal secondary endpoints:

\begin{enumerate}
    \item WHO 8-point ordinal outcome scale at day 28 post randomisation: (1) not hospitalised and no limitations on activities, (2) not hospitalised, limitations on activities, (3) hospitalised, not requiring supplemental oxygen and no longer requiring ongoing medical care, (4) hospitalised, not requiring supplemental oxygen but requiring ongoing medical care, (5) hospitalised, requiring supplemental oxygen, (6) hospitalised, on non-invasive ventilation or high flow oxygen devices, (7) hospitalised, on invasive mechanical ventilation or ECMO, and (8) death.
    \item Modified Medical Research Council (mMRC) breathlessness scale (only asked among those diagnosed with COVID-19), a 5-point ordinal scale: (1) `I only get breathless with strenuous exercise', (2) `I get short of breath when hurrying on level ground or walking up a slight hill', (3) `On level ground, I walk slower than people of the same age because of breathlessness, or I have to stop for breath when walking at my own pace on the level', (4) `I stop for breath after walking about 100 yards or after a few minutes on level ground', and (5) `I am too breathless to leave the house or I am breathless when dressing or undressing'.
    \item Days alive and free of hospital by 28 days post randomisation: a 29-point ordinal outcome calculated as 28 minus the number of days spent in hospital, where patients dying within 28 days were assigned zero free days. 
    \item Days alive and free of ventilation by 28 days post randomisation: a 29-point ordinal outcome calculated as 28 minus the number of days free of invasive or non-invasive ventilation, where patients dying within 28 days were assigned zero free days. 
\end{enumerate}

The same prior distributions used in the simulation study were applied in the analysis of the anticoagulation domain of ASCOT using complete cases only \cite{mcquilten2023anticoagulation}. For simplicity we focused on comparing only the low vs intermediate dose with LMWH and present the results for each of the four secondary ordinal endpoints. Supplementary Material 4 provides an overview of the distribution across categories by intervention group for each ordinal endpoint, which were largely skewed. 

When the six priors for the treatment effect were applied to the analysis of the WHO scale, all methods estimated the proportional log-OR of the ordinal scale to be smaller than zero indicating the low-dose provided better odds of a more favourable outcome (Figure 4). All approaches resulted in a similar point estimate and large uncertainty irrespective of the prior (Figure 3). Similar estimated treatment effects were observed when the implicit prior on the cut-points is varied for all of these outcomes. Supplementary Material 4 indicates similar results across all other outcomes.

\begin{figure}[!htbp]
\centering
\caption{Estimated treatment effect of proportional log-ORs for the WHO eight-point scale in the Australasian COVID-19 Trial (ASCOT) obtained under the six prior specifications for treatment effect}
\small\textsuperscript{OR = odds ratio, SD = standard deviation}

\advance\leftskip-3cm
\advance\rightskip-3cm
\includegraphics[width=1\textwidth]{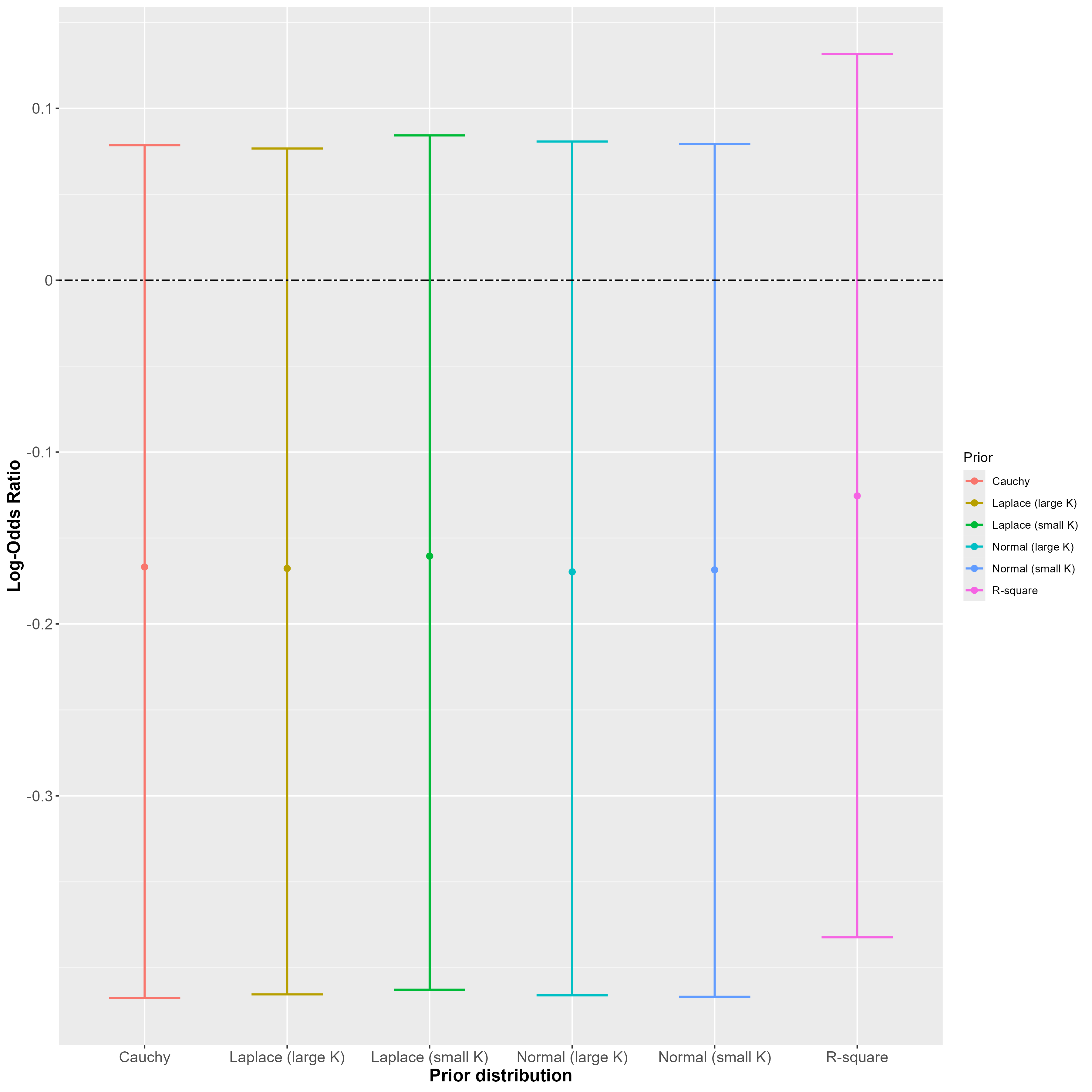}
\end{figure}

\newpage 

\section*{Discussion}

This study examined the impact of different non-informative prior distributions on parameters of a Bayesian PO model have on estimation and operating characteristics in the context of a fixed and an adaptive trial with an early stopping rule. When varying the prior for the treatment effect, results indicate that bias was substantial when the control arm probabilities are right-skewed for all priors apart from the R-squared prior specification, especially with increasing categories in the ordinal outcome and smaller sample size. If it is expected that participants will fall at both ends of the ordinal scale, then specifying a Normal, Cauchy or Laplace prior distribution for the treatment effect would be appropriate. There was not much variation in the results when a Laplace or Normal prior distribution was used for the treatment effect or with different values for the standard deviation for these priors. This implies that in fact, specifying a reasonably small or large variance term will still result in a rather `non-informative' prior distribution that does not impact the treatment effect in the scenarios considered in this study. This study also found that when altering the implicit prior on the model cut-points while keeping the prior for the treatment effect fixed, decreasing the concentration parameters for a Dirichlet prior can reduce bias in treatment effect estimation for right-skewed control arm probabilities, with minimal bias when using a Normal distribution for cut-points (except for small sample sizes and 30 categories), and results in a higher proportion of trials to stop early. 

The results from our study also show that it is more likely that divergent transitions will occur when a Dirichlet prior with concentration parameters close to zero is used for the control group probabilities which, although can attempt to be mitigated by increasing the target average acceptance probability rate, may still exist in some circumstances. Divergent transitions implies that there were issues in the sampler exploring the posterior distribution of the target parameter, which can lead to biased inference. Divergent transitions were more likely to occur when the sample size is small and with a larger number of categories. The implications of this is that careful consideration should be made for selecting an appropriate Dirichlet prior, particularly when it is known that the interim analysis will occur at a small sample size or the ordinal outcome has a large number of categories. 

Symmetric priors like the Normal, Laplace or Cauchy distributions assume that the treatment effect could be in either direction. This may be unjustified if a large proportion of participants are expected to be at one end of the scale (and therefore there is data sparsity in the remaining categories) which may explain the bias with these approaches when the controlled probabilities are skewed.  In contrast, the R-squared prior allows for the fact that the variation in the outcome might not be due to the treatment effect alone, but could also be explained by other factors such as the distribution of outcomes in the control arm. This could be more consistent with the skewed distribution of outcomes and hence may explain the minimal bias with this approach. When varying the prior for the control arm probabilities, the uniform Dirichlet prior (which reflects that there is equal probability in each category of the control arm) may lead to poor estimation of the control arm’s distribution in the PO model, particularly when the control probabilities are skewed, resulting in an underestimation of the treatment effect. Trials that incorporated early stopping rules exhibited larger bias in the treatment effect for some of the priors than the fixed design, particularly for U-shaped control probabilities. This aligns with the expectations that trials that stop early are more likely due to random highs \cite{walter2019randomised}. Adaptive designs that used the R-squared prior specification had the highest early stopping rates when the control data were skewed, particularly when the ordinal outcome had many categories. 

We also found that decreasing the concentration parameters for a Dirichlet prior can reduce bias in treatment effect estimation for right-skewed control arm probabilities. Reducing the concentration parameters allow the Dirichlet prior to become more diffuse, meaning it gives less weight to a prior belief that the categories should be equally distributed and uniform. This therefore gives the model more freedom to adjust based on the actual data. When the control arm probabilities were U-shaped, the choice of prior for the control group probabilities had little impact on the estimation of the treatment effect, likely because the prior distributions are symmetric about the middle category and there is reasonable probability of falling in each category. We found that using a Dirichlet prior for the control group probabilities with a concentration parameter closer to zero resulted in a trial that was more likely to stop early and/or declare superiority, even when there was no association between intervention and outcome. This is likely because the lower value of the concentration parameter leads to more uncertainty in the prior belief about the outcome. This flexibility makes the model more sensitive to random fluctuations in the data, increasing the chances of detecting a treatment effect even in the absence of a true effect, leading to premature stopping of the trial. Another possible explanation is that the priors for the model parameters for the treatment effect and model `intercepts' have been specified independently from each other \textit{a priori} (and are therefore approximately independent in the posterior). However, in our study, we used standard treatment coding (assigning 0 and 1 for the treatment indicator) which does not ensure that these parameters are uncorrelated in the model. An alternative would be to center the treatment contrast specification, that is, to code the treatment variable as $\pm \frac{1}{2}$ \cite{kraemer2004centring}, to ensure the treatment coefficient is approximately uncorrelated with the model `intercepts' which may alleviate the observed bias. Although not the focus of this study, future research could explore centering of treatment covariates in more detail. 

The results of this study has implications for practice. It is common for trials (whether fixed or adaptive) to specify a Normal prior distribution for the treatment effect and an implicit Dirichlet prior on the model cut-points in a Bayesian PO model \cite{selman2024statistical} using $\boldsymbol{\alpha = 1}$. However, the results from this study suggest that the most appropriate priors may depend on the distribution of participants in the control arm. Use of pilot data or similar studies would therefore be useful prior to data collection in determining the distribution of patients across the categories in order to guide the selection of priors. For example, if it is expected that control arm patients are more likely to fall in the first few categories of the ordinal outcome, then an R-squared prior specification on the treatment effect might be more appropriate, particularly if the ordinal outcome has a large number of categories. If little is known about the distribution of the outcome in advance, then pre-specified sensitivity analyses could be conducted using various non-informative priors to assess the robustness of the results to the priors used at the final analysis. For example, if a diffuse Normal prior is used for the treatment effect, which is common in practice, then decreasing the value of the concentration parameters of the Dirichlet prior closer to zero or using a Normal prior for the cut-points may reduce bias and increase the likelihood of stopping the trial early when the control arm probabilities are skewed. We acknowledge that an exploration of the different priors would not be as useful for an interim analysis if one prior resulted in a threshold being reached to stop but another prior did not. 

Our study has several strengths. The simulation study explores a number of realistic scenarios (108 in total). We have considered both a fixed and an adaptive design, and considered a range of prior distributions that could be used. We have also applied the same approaches to a case study with four different ordinal outcomes. However, this study does have its limitations. First, there are many other possible control arm distributions and adaptive designs that could be explored where the different non-informative priors could lead to different results. Second, we did not explore the impact that the specification of the priors may have in a multi-arm trial or in an analysis that adjusts for covariates in the model, such as randomisation strata, although we do not expect the latter to affect the results. Third, we only considered a limited set of commonly used priors and acknowledge that there are other alternatives that may warrant investigation. Finally, we did not consider the impact that departure from PO has on the results as this may not be a realistic assumption to make in practice. This should be investigated in future studies.  

In summary, there are numerous possible non-informative prior distributions that can be specified for the Bayesian PO model, and careful specification of these should be considered when conducting such an analysis. Importantly, the most appropriate prior appears to depend on the distribution of the control arm probabilities. We recommend the use of the R-squared prior specification when the control arm is skewed, and a Dirichlet prior with concentration parameters close to zero in conjunction with a Normal prior for the treatment effect if the analysis occurs at a reasonable sample size (e.g. $n = 500$), otherwise concentration parameters set to one would be appropriate to avoid divergent transitions. Choosing an appropriate non-informative prior will result in better inference and therefore trial conclusions and adaptive decision-making.

\newpage 


\begin{backmatter}
\section*{Declarations}
\subsection*{Acknowledgements}
The authors would like to thank the Australian COVID-19 (ASCOT) trial team for providing access and use of the ASCOT trial data to illustrate the methods discussed in this paper, and to ASCOT investigator and statistical team member Tom Snelling for providing useful insights when shaping the manuscript. 

\subsection*{Funding}
This work forms part of Chris Selman’s PhD, which is supported by the Research Training Program Scholarship, administered by the Australian Commonwealth Government and The University of Melbourne, Australia. Chris Selman's PhD was also supported by a Centre of Research Excellence grant from the National Health and Medical Research Council of Australia ID 1171422, to the AusTriM Research Network. Research at the Murdoch Children’s Research Institute is supported by the Victorian Government’s Operational Infrastructure Support Program. This work was supported by the Australian National Health and Medical Research Council (NHMRC) Centre for Research Excellence grants to the Victorian Centre for Biostatistics (ID 1035261) and the Australian Trials Methodology Research Network (ID 1171422), including through seed funding awarded to Robert Mahar. Katherine Lee is funded by an NHMRC investigator grant (ID 2017498). Michael Dymock is supported by a NHMRC Postgraduate Research Award (APP2022557). The funding bodies played no role in the study conception, design, data collection, data analysis, data interpretation or writing of the report.

\subsection*{Consent for Publication}
Not applicable.

\subsection*{Availability of data and materials}
The code and datasets generated for this simulation study are available on Github \cite{SelmanGit2025}: \url{https://github.com/chrisselman/noninfo_pomodel}.

Data used for the illustrative example within this paper are available upon reasonable request from the ASCOT study team.

\subsection*{Abbreviations}
ASCOT: Australasian COVID-19 Trial, ESS: Effective Sample Size, HMC: Hamiltonian Monte Carlo, LMWH: Low-Molecular-Weight Heparin, RCT: randomised controlled trial, OR: odds ratio, PO: proportional odds, MCSE: Monte Carlo standard error, mMRC: Modified Medical Research Council, WHO: World Health Organisation. 

\subsection*{Ethics approval and consent to participate}
Use of the ASCOT data was approved by both the ASCOT Global Trial Steering Committee and the MCRI Change Advisory Board.

\subsection*{Registration details}
The ASCOT trial is registered on the Australian New Zealand Clinical Trials Registry (ANZCTR) with registration number ACTRN12620000445976 on the 6 April 2020.

\subsection*{Competing interests}
The authors declare that they have no competing interests.

\subsection*{Authors' contributions}
CJS, KJL, RKM conceived the study. CJS conducted the simulation study and data analysis and wrote the first draft of the manuscript, with regular input from KJL and RKM. All authors contributed to the design of the study, critical revision of the manuscript, and take responsibility for its content.


\bibliographystyle{bmc-mathphys} 
\bibliography{bmc_article}      


\begin{thebibliography}{37}
\ifx \bisbn   \undefined \def \bisbn  #1{ISBN #1}\fi
\ifx \binits  \undefined \def \binits#1{#1}\fi
\ifx \bauthor  \undefined \def \bauthor#1{#1}\fi
\ifx \batitle  \undefined \def \batitle#1{#1}\fi
\ifx \bjtitle  \undefined \def \bjtitle#1{#1}\fi
\ifx \bvolume  \undefined \def \bvolume#1{\textbf{#1}}\fi
\ifx \byear  \undefined \def \byear#1{#1}\fi
\ifx \bissue  \undefined \def \bissue#1{#1}\fi
\ifx \bfpage  \undefined \def \bfpage#1{#1}\fi
\ifx \blpage  \undefined \def \blpage #1{#1}\fi
\ifx \burl  \undefined \def \burl#1{\textsf{#1}}\fi
\ifx \doiurl  \undefined \def \doiurl#1{\textsf{#1}}\fi
\ifx \betal  \undefined \def \betal{\textit{et al.}}\fi
\ifx \binstitute  \undefined \def \binstitute#1{#1}\fi
\ifx \binstitutionaled  \undefined \def \binstitutionaled#1{#1}\fi
\ifx \bctitle  \undefined \def \bctitle#1{#1}\fi
\ifx \beditor  \undefined \def \beditor#1{#1}\fi
\ifx \bpublisher  \undefined \def \bpublisher#1{#1}\fi
\ifx \bbtitle  \undefined \def \bbtitle#1{#1}\fi
\ifx \bedition  \undefined \def \bedition#1{#1}\fi
\ifx \bseriesno  \undefined \def \bseriesno#1{#1}\fi
\ifx \blocation  \undefined \def \blocation#1{#1}\fi
\ifx \bsertitle  \undefined \def \bsertitle#1{#1}\fi
\ifx \bsnm \undefined \def \bsnm#1{#1}\fi
\ifx \bsuffix \undefined \def \bsuffix#1{#1}\fi
\ifx \bparticle \undefined \def \bparticle#1{#1}\fi
\ifx \barticle \undefined \def \barticle#1{#1}\fi
\ifx \bconfdate \undefined \def \bconfdate #1{#1}\fi
\ifx \botherref \undefined \def \botherref #1{#1}\fi
\ifx \url \undefined \def \url#1{\textsf{#1}}\fi
\ifx \bchapter \undefined \def \bchapter#1{#1}\fi
\ifx \bbook \undefined \def \bbook#1{#1}\fi
\ifx \bcomment \undefined \def \bcomment#1{#1}\fi
\ifx \oauthor \undefined \def \oauthor#1{#1}\fi
\ifx \citeauthoryear \undefined \def \citeauthoryear#1{#1}\fi
\ifx \endbibitem  \undefined \def \endbibitem {}\fi
\ifx \bconflocation  \undefined \def \bconflocation#1{#1}\fi
\ifx \arxivurl  \undefined \def \arxivurl#1{\textsf{#1}}\fi
\csname PreBibitemsHook\endcsname

\bibitem{selman2024statistical}
\begin{barticle}
\bauthor{\bsnm{Selman}, \binits{C.J.}},
\bauthor{\bsnm{Lee}, \binits{K.J.}},
\bauthor{\bsnm{Ferguson}, \binits{K.N.}},
\bauthor{\bsnm{Whitehead}, \binits{C.L.}},
\bauthor{\bsnm{Manley}, \binits{B.J.}},
\bauthor{\bsnm{Mahar}, \binits{R.K.}}:
\batitle{Statistical analyses of ordinal outcomes in randomised controlled trials: a scoping review}.
\bjtitle{Trials}
\bvolume{25}(\bissue{1}),
\bfpage{1}--\blpage{18}
(\byear{2024})
\end{barticle}
\endbibitem

\bibitem{bonita1988recovery}
\begin{barticle}
\bauthor{\bsnm{Bonita}, \binits{R.}},
\bauthor{\bsnm{Beaglehole}, \binits{R.}}:
\batitle{Recovery of motor function after stroke.}
\bjtitle{Stroke}
\bvolume{19}(\bissue{12}),
\bfpage{1497}--\blpage{1500}
(\byear{1988})
\end{barticle}
\endbibitem

\bibitem{mathioudakis2020outcomes}
\begin{barticle}
\bauthor{\bsnm{Mathioudakis}, \binits{A.G.}},
\bauthor{\bsnm{Fally}, \binits{M.}},
\bauthor{\bsnm{Hashad}, \binits{R.}},
\bauthor{\bsnm{Kouta}, \binits{A.}},
\bauthor{\bsnm{Hadi}, \binits{A.S.}},
\bauthor{\bsnm{Knight}, \binits{S.B.}},
\bauthor{\bsnm{Bakerly}, \binits{N.D.}},
\bauthor{\bsnm{Singh}, \binits{D.}},
\bauthor{\bsnm{Williamson}, \binits{P.R.}},
\bauthor{\bsnm{Felton}, \binits{T.}}, \betal:
\batitle{Outcomes evaluated in controlled clinical trials on the management of {COVID}-19: a methodological systematic review}.
\bjtitle{Life}
\bvolume{10}(\bissue{12}),
\bfpage{350}
(\byear{2020})
\end{barticle}
\endbibitem

\bibitem{walker1967estimation}
\begin{barticle}
\bauthor{\bsnm{Walker}, \binits{S.H.}},
\bauthor{\bsnm{Duncan}, \binits{D.B.}}:
\batitle{Estimation of the probability of an event as a function of several independent variables}.
\bjtitle{Biometrika}
\bvolume{54}(\bissue{1-2}),
\bfpage{167}--\blpage{179}
(\byear{1967})
\end{barticle}
\endbibitem

\bibitem{mccullagh1980regression}
\begin{barticle}
\bauthor{\bsnm{McCullagh}, \binits{P.}}:
\batitle{Regression models for ordinal data}.
\bjtitle{Journal of the Royal Statistical Society: Series B (Methodological)}
\bvolume{42}(\bissue{2}),
\bfpage{109}--\blpage{127}
(\byear{1980})
\end{barticle}
\endbibitem

\bibitem{harrell2015regression}
\begin{bbook}
\bauthor{\bsnm{Harrell}, \binits{F.E.}}, \betal:
\bbtitle{Regression Modeling Strategies: with Applications to Linear Models, Logistic and Ordinal Regression, and Survival Analysis}
vol. \bseriesno{3}.
\bpublisher{Springer}, \blocation{???}
(\byear{2015})
\end{bbook}
\endbibitem

\bibitem{stallard2020efficient}
\begin{barticle}
\bauthor{\bsnm{Stallard}, \binits{N.}},
\bauthor{\bsnm{Hampson}, \binits{L.}},
\bauthor{\bsnm{Benda}, \binits{N.}},
\bauthor{\bsnm{Brannath}, \binits{W.}},
\bauthor{\bsnm{Burnett}, \binits{T.}},
\bauthor{\bsnm{Friede}, \binits{T.}},
\bauthor{\bsnm{Kimani}, \binits{P.K.}},
\bauthor{\bsnm{Koenig}, \binits{F.}},
\bauthor{\bsnm{Krisam}, \binits{J.}},
\bauthor{\bsnm{Mozgunov}, \binits{P.}}, \betal:
\batitle{Efficient adaptive designs for clinical trials of interventions for covid-19}.
\bjtitle{Statistics in Biopharmaceutical Research}
\bvolume{12}(\bissue{4}),
\bfpage{483}--\blpage{497}
(\byear{2020})
\end{barticle}
\endbibitem

\bibitem{perkins2020recovery}
\begin{barticle}
\bauthor{\bsnm{Perkins}, \binits{G.D.}},
\bauthor{\bsnm{Couper}, \binits{K.}},
\bauthor{\bsnm{Connolly}, \binits{B.}},
\bauthor{\bsnm{Baillie}, \binits{J.K.}},
\bauthor{\bsnm{Bradley}, \binits{J.M.}},
\bauthor{\bsnm{Dark}, \binits{P.}},
\bauthor{\bsnm{De~Soyza}, \binits{A.}},
\bauthor{\bsnm{Gorman}, \binits{E.}},
\bauthor{\bsnm{Gray}, \binits{A.}},
\bauthor{\bsnm{Hamilton}, \binits{L.}}, \betal:
\batitle{Recovery-respiratory support: respiratory strategies for patients with suspected or proven covid-19 respiratory failure; continuous positive airway pressure, high-flow nasal oxygen, and standard care: a structured summary of a study protocol for a randomised controlled trial}.
\bjtitle{Trials}
\bvolume{21}(\bissue{1}),
\bfpage{1}--\blpage{3}
(\byear{2020})
\end{barticle}
\endbibitem

\bibitem{florescu2022effect}
\begin{barticle}
\bauthor{\bsnm{Florescu}, \binits{S.}},
\bauthor{\bsnm{Stanciu}, \binits{D.}},
\bauthor{\bsnm{Zaharia}, \binits{M.}},
\bauthor{\bsnm{Kosa}, \binits{A.}},
\bauthor{\bsnm{Codreanu}, \binits{D.}},
\bauthor{\bsnm{Kidwai}, \binits{A.}},
\bauthor{\bsnm{Masood}, \binits{S.}},
\bauthor{\bsnm{Kaye}, \binits{C.}},
\bauthor{\bsnm{Coutts}, \binits{A.}},
\bauthor{\bsnm{MacKay}, \binits{L.}}, \betal:
\batitle{Effect of antiplatelet therapy on survival and organ support--free days in critically ill patients with covid-19: a randomized clinical trial}.
\bjtitle{Jama}
\bvolume{327}(\bissue{13}),
\bfpage{1247}--\blpage{1259}
(\byear{2022})
\end{barticle}
\endbibitem

\bibitem{kairalla2012adaptive}
\begin{barticle}
\bauthor{\bsnm{Kairalla}, \binits{J.A.}},
\bauthor{\bsnm{Coffey}, \binits{C.S.}},
\bauthor{\bsnm{Thomann}, \binits{M.A.}},
\bauthor{\bsnm{Muller}, \binits{K.E.}}:
\batitle{Adaptive trial designs: a review of barriers and opportunities}.
\bjtitle{Trials}
\bvolume{13},
\bfpage{1}--\blpage{9}
(\byear{2012})
\end{barticle}
\endbibitem

\bibitem{mcmurray2014angiotensin}
\begin{barticle}
\bauthor{\bsnm{McMurray}, \binits{J.J.}},
\bauthor{\bsnm{Packer}, \binits{M.}},
\bauthor{\bsnm{Desai}, \binits{A.S.}},
\bauthor{\bsnm{Gong}, \binits{J.}},
\bauthor{\bsnm{Lefkowitz}, \binits{M.P.}},
\bauthor{\bsnm{Rizkala}, \binits{A.R.}},
\bauthor{\bsnm{Rouleau}, \binits{J.L.}},
\bauthor{\bsnm{Shi}, \binits{V.C.}},
\bauthor{\bsnm{Solomon}, \binits{S.D.}},
\bauthor{\bsnm{Swedberg}, \binits{K.}}, \betal:
\batitle{Angiotensin--neprilysin inhibition versus enalapril in heart failure}.
\bjtitle{N Engl J Med}
\bvolume{371},
\bfpage{993}--\blpage{1004}
(\byear{2014})
\end{barticle}
\endbibitem

\bibitem{berry2006bayesian}
\begin{barticle}
\bauthor{\bsnm{Berry}, \binits{D.A.}}:
\batitle{Bayesian clinical trials}.
\bjtitle{Nature reviews Drug discovery}
\bvolume{5}(\bissue{1}),
\bfpage{27}--\blpage{36}
(\byear{2006})
\end{barticle}
\endbibitem

\bibitem{bast2010holland}
\begin{bbook}
\bauthor{\bsnm{Bast~Jr}, \binits{R.C.}},
\bauthor{\bsnm{Holland}, \binits{J.F.}}:
\bbtitle{Holland-Frei Cancer Medicine 8}.
\bpublisher{PMPH-USA}, \blocation{???}
(\byear{2010})
\end{bbook}
\endbibitem

\bibitem{berry2005introduction}
\begin{botherref}
\oauthor{\bsnm{Berry}, \binits{D.A.}}:
Introduction to Bayesian methods III: use and interpretation of Bayesian tools in design and analysis.
Sage Publications Sage CA: Thousand Oaks, CA
(2005)
\end{botherref}
\endbibitem

\bibitem{gelman2013philosophy}
\begin{barticle}
\bauthor{\bsnm{Gelman}, \binits{A.}},
\bauthor{\bsnm{Shalizi}, \binits{C.R.}}:
\batitle{Philosophy and the practice of bayesian statistics}.
\bjtitle{British Journal of Mathematical and Statistical Psychology}
\bvolume{66}(\bissue{1}),
\bfpage{8}--\blpage{38}
(\byear{2013})
\end{barticle}
\endbibitem

\bibitem{mckinley2015bayesian}
\begin{botherref}
\oauthor{\bsnm{McKinley}, \binits{T.J.}},
\oauthor{\bsnm{Morters}, \binits{M.}},
\oauthor{\bsnm{Wood}, \binits{J.L.}}:
Bayesian model choice in cumulative link ordinal regression models
(2015)
\end{botherref}
\endbibitem

\bibitem{james2021bayesian}
\begin{botherref}
\oauthor{\bsnm{James}, \binits{N.T.}},
\oauthor{\bsnm{Harrell~Jr}, \binits{F.E.}},
\oauthor{\bsnm{Shepherd}, \binits{B.E.}}:
Bayesian cumulative probability models for continuous and mixed outcomes.
arXiv preprint arXiv:2102.00330
(2021)
\end{botherref}
\endbibitem

\bibitem{mcquilten2023anticoagulation}
\begin{barticle}
\bauthor{\bsnm{McQuilten}, \binits{Z.K.}},
\bauthor{\bsnm{Venkatesh}, \binits{B.}},
\bauthor{\bsnm{Jha}, \binits{V.}},
\bauthor{\bsnm{Roberts}, \binits{J.}},
\bauthor{\bsnm{Morpeth}, \binits{S.C.}},
\bauthor{\bsnm{Totterdell}, \binits{J.A.}},
\bauthor{\bsnm{McPhee}, \binits{G.M.}},
\bauthor{\bsnm{Abraham}, \binits{J.}},
\bauthor{\bsnm{Bam}, \binits{N.}},
\bauthor{\bsnm{Bandara}, \binits{M.}}, \betal:
\batitle{Anticoagulation strategies in non--critically ill patients with covid-19}.
\bjtitle{NEJM Evidence}
\bvolume{2}(\bissue{2}),
\bfpage{2200293}
(\byear{2023})
\end{barticle}
\endbibitem

\bibitem{gelman1995bayesian}
\begin{bbook}
\bauthor{\bsnm{Gelman}, \binits{A.}},
\bauthor{\bsnm{Carlin}, \binits{J.B.}},
\bauthor{\bsnm{Stern}, \binits{H.S.}},
\bauthor{\bsnm{Rubin}, \binits{D.B.}}:
\bbtitle{Bayesian Data Analysis}.
\bpublisher{Chapman and Hall/CRC}, \blocation{???}
(\byear{2014})
\end{bbook}
\endbibitem

\bibitem{gelman2008weakly}
\begin{botherref}
\oauthor{\bsnm{Gelman}, \binits{A.}},
\oauthor{\bsnm{Jakulin}, \binits{A.}},
\oauthor{\bsnm{Pittau}, \binits{M.G.}},
\oauthor{\bsnm{Su}, \binits{Y.-S.}}:
A weakly informative default prior distribution for logistic and other regression models
(2008)
\end{botherref}
\endbibitem

\bibitem{sarma2020prior}
\begin{bchapter}
\bauthor{\bsnm{Sarma}, \binits{A.}},
\bauthor{\bsnm{Kay}, \binits{M.}}:
\bctitle{Prior setting in practice: Strategies and rationales used in choosing prior distributions for bayesian analysis}.
In: \bbtitle{Proceedings of the 2020 Chi Conference on Human Factors in Computing Systems},
pp. \bfpage{1}--\blpage{12}
(\byear{2020})
\end{bchapter}
\endbibitem

\bibitem{rochfordprior}
\begin{botherref}
\oauthor{\bsnm{Rochford}, \binits{A.}}:
Prior Distributions for Bayesian Regression Using PyMC.
\url{https://austinrochford.com/posts/2013-09-02-prior-distributions-for-bayesian-regression-using-pymc.html}.
Accessed: 2025-01-20
(2013)
\end{botherref}
\endbibitem

\bibitem{rsquare1}
\begin{botherref}
\oauthor{\bsnm{Gabry}, \binits{J.}},
\oauthor{\bsnm{Goodrich}, \binits{B.}}:
Estimating Regularized Linear Models with rstanarm.
\url{https://cran.r-project.org/web/packages/rstanarm/vignettes/lm.html#priors}.
Accessed: 2025-01-20
(2024)
\end{botherref}
\endbibitem

\bibitem{tu2014dirichlet}
\begin{botherref}
\oauthor{\bsnm{Tu}, \binits{S.}}:
The dirichlet-multinomial and dirichlet-categorical models for bayesian inference.
Computer Science Division, UC Berkeley
\textbf{2}
(2014)
\end{botherref}
\endbibitem

\bibitem{hill2020efficacy}
\begin{barticle}
\bauthor{\bsnm{Hill}, \binits{M.D.}},
\bauthor{\bsnm{Goyal}, \binits{M.}},
\bauthor{\bsnm{Menon}, \binits{B.K.}},
\bauthor{\bsnm{Nogueira}, \binits{R.G.}},
\bauthor{\bsnm{McTaggart}, \binits{R.A.}},
\bauthor{\bsnm{Demchuk}, \binits{A.M.}},
\bauthor{\bsnm{Poppe}, \binits{A.Y.}},
\bauthor{\bsnm{Buck}, \binits{B.H.}},
\bauthor{\bsnm{Field}, \binits{T.S.}},
\bauthor{\bsnm{Dowlatshahi}, \binits{D.}}, \betal:
\batitle{Efficacy and safety of nerinetide for the treatment of acute ischaemic stroke (escape-na1): a multicentre, double-blind, randomised controlled trial}.
\bjtitle{The Lancet}
\bvolume{395}(\bissue{10227}),
\bfpage{878}--\blpage{887}
(\byear{2020})
\end{barticle}
\endbibitem

\bibitem{rstudio}
\begin{bbook}
\bauthor{\bsnm{{R Core Team}}}:
\bbtitle{R: A Language and Environment for Statistical Computing}.
\bpublisher{R Foundation for Statistical Computing},
\blocation{Vienna, Austria}
(\byear{2021}).
\bcomment{R Foundation for Statistical Computing}.
\burl{https://www.R-project.org/}
\end{bbook}
\endbibitem

\bibitem{carpenter2017stan}
\begin{botherref}
\oauthor{\bsnm{Carpenter}, \binits{B.}},
\oauthor{\bsnm{Gelman}, \binits{A.}},
\oauthor{\bsnm{Hoffman}, \binits{M.D.}},
\oauthor{\bsnm{Lee}, \binits{D.}},
\oauthor{\bsnm{Goodrich}, \binits{B.}},
\oauthor{\bsnm{Betancourt}, \binits{M.}},
\oauthor{\bsnm{Brubaker}, \binits{M.}},
\oauthor{\bsnm{Guo}, \binits{J.}},
\oauthor{\bsnm{Li}, \binits{P.}},
\oauthor{\bsnm{Riddell}, \binits{A.}}:
Stan: A probabilistic programming language.
Journal of statistical software
\textbf{76}(1)
(2017)
\end{botherref}
\endbibitem

\bibitem{posteriorpackage}
\begin{botherref}
\oauthor{\bsnm{Bürkner}, \binits{P.-C.}},
\oauthor{\bsnm{Gabry}, \binits{J.}},
\oauthor{\bsnm{Kay}, \binits{M.}},
\oauthor{\bsnm{Vehtari}, \binits{A.}}:
posterior: Tools for Working with Posterior Distributions.
R package version 1.6.0
(2024).
\url{https://mc-stan.org/posterior/}
\end{botherref}
\endbibitem

\bibitem{Goodrich2020}
\begin{botherref}
\oauthor{\bsnm{Goodrich}, \binits{B.}},
\oauthor{\bsnm{Gabry}, \binits{J.}},
\oauthor{\bsnm{Ali}, \binits{I.}},
\oauthor{\bsnm{Brilleman}, \binits{S.}}:
rstanarm: {Bayesian} applied regression modeling via {Stan}
(2020).
\url{https://mc-stan.org/rstanarm}
\end{botherref}
\endbibitem

\bibitem{kelter2023bayesian}
\begin{botherref}
\oauthor{\bsnm{Kelter}, \binits{R.}}:
The bayesian simulation study (basis) framework for simulation studies in statistical and methodological research.
Biometrical Journal,
2200095
(2023)
\end{botherref}
\endbibitem

\bibitem{koehler2009assessment}
\begin{barticle}
\bauthor{\bsnm{Koehler}, \binits{E.}},
\bauthor{\bsnm{Brown}, \binits{E.}},
\bauthor{\bsnm{Haneuse}, \binits{S.J.-P.}}:
\batitle{On the assessment of monte carlo error in simulation-based statistical analyses}.
\bjtitle{The American Statistician}
\bvolume{63}(\bissue{2}),
\bfpage{155}--\blpage{162}
(\byear{2009})
\end{barticle}
\endbibitem

\bibitem{morris2019using}
\begin{barticle}
\bauthor{\bsnm{Morris}, \binits{T.P.}},
\bauthor{\bsnm{White}, \binits{I.R.}},
\bauthor{\bsnm{Crowther}, \binits{M.J.}}:
\batitle{Using simulation studies to evaluate statistical methods}.
\bjtitle{Statistics in medicine}
\bvolume{38}(\bissue{11}),
\bfpage{2074}--\blpage{2102}
(\byear{2019})
\end{barticle}
\endbibitem

\bibitem{goodrich2020rstanarm}
\begin{botherref}
\oauthor{\bsnm{Goodrich}, \binits{B.}},
\oauthor{\bsnm{Gabry}, \binits{J.}},
\oauthor{\bsnm{Ali}, \binits{I.}},
\oauthor{\bsnm{Brilleman}, \binits{S.}}:
rstanarm: Bayesian applied regression modeling via stan.
R package version
\textbf{2}(1)
(2020)
\end{botherref}
\endbibitem

\bibitem{vehtari2021rank}
\begin{barticle}
\bauthor{\bsnm{Vehtari}, \binits{A.}},
\bauthor{\bsnm{Gelman}, \binits{A.}},
\bauthor{\bsnm{Simpson}, \binits{D.}},
\bauthor{\bsnm{Carpenter}, \binits{B.}},
\bauthor{\bsnm{B{\"u}rkner}, \binits{P.-C.}}:
\batitle{Rank-normalization, folding, and localization: An improved r ̂ for assessing convergence of mcmc (with discussion)}.
\bjtitle{Bayesian analysis}
\bvolume{16}(\bissue{2}),
\bfpage{667}--\blpage{718}
(\byear{2021})
\end{barticle}
\endbibitem

\bibitem{walter2019randomised}
\begin{barticle}
\bauthor{\bsnm{Walter}, \binits{S.}},
\bauthor{\bsnm{Guyatt}, \binits{G.}},
\bauthor{\bsnm{Bassler}, \binits{D.}},
\bauthor{\bsnm{Briel}, \binits{M.}},
\bauthor{\bsnm{Ramsay}, \binits{T.}},
\bauthor{\bsnm{Han}, \binits{H.}}:
\batitle{Randomised trials with provision for early stopping for benefit (or harm): the impact on the estimated treatment effect}.
\bjtitle{Statistics in medicine}
\bvolume{38}(\bissue{14}),
\bfpage{2524}--\blpage{2543}
(\byear{2019})
\end{barticle}
\endbibitem

\bibitem{kraemer2004centring}
\begin{barticle}
\bauthor{\bsnm{Kraemer}, \binits{H.C.}},
\bauthor{\bsnm{Blasey}, \binits{C.M.}}:
\batitle{Centring in regression analyses: a strategy to prevent errors in statistical inference}.
\bjtitle{International journal of methods in psychiatric research}
\bvolume{13}(\bissue{3}),
\bfpage{141}--\blpage{151}
(\byear{2004})
\end{barticle}
\endbibitem

\bibitem{SelmanGit2025}
\begin{botherref}
\oauthor{\bsnm{Selman}, \binits{C.J.}}:
Simulation Study Code.
GitHub
(2025)
\end{botherref}
\endbibitem

\end{thebibliography}

\newcommand{\BMCxmlcomment}[1]{}

\BMCxmlcomment{

<refgrp>

<bibl id="B1">
  <title><p>Statistical analyses of ordinal outcomes in randomised controlled trials: a scoping review</p></title>
  <aug>
    <au><snm>Selman</snm><fnm>CJ</fnm></au>
    <au><snm>Lee</snm><fnm>KJ</fnm></au>
    <au><snm>Ferguson</snm><fnm>KN</fnm></au>
    <au><snm>Whitehead</snm><fnm>CL</fnm></au>
    <au><snm>Manley</snm><fnm>BJ</fnm></au>
    <au><snm>Mahar</snm><fnm>RK</fnm></au>
  </aug>
  <source>Trials</source>
  <publisher>Springer</publisher>
  <pubdate>2024</pubdate>
  <volume>25</volume>
  <issue>1</issue>
  <fpage>1</fpage>
  <lpage>-18</lpage>
</bibl>

<bibl id="B2">
  <title><p>Recovery of motor function after stroke.</p></title>
  <aug>
    <au><snm>Bonita</snm><fnm>R</fnm></au>
    <au><snm>Beaglehole</snm><fnm>R</fnm></au>
  </aug>
  <source>Stroke</source>
  <publisher>Am Heart Assoc</publisher>
  <pubdate>1988</pubdate>
  <volume>19</volume>
  <issue>12</issue>
  <fpage>1497</fpage>
  <lpage>-1500</lpage>
</bibl>

<bibl id="B3">
  <title><p>Outcomes evaluated in controlled clinical trials on the management of {COVID}-19: a methodological systematic review</p></title>
  <aug>
    <au><snm>Mathioudakis</snm><fnm>AG</fnm></au>
    <au><snm>Fally</snm><fnm>M</fnm></au>
    <au><snm>Hashad</snm><fnm>R</fnm></au>
    <au><snm>Kouta</snm><fnm>A</fnm></au>
    <au><snm>Hadi</snm><fnm>AS</fnm></au>
    <au><snm>Knight</snm><fnm>SB</fnm></au>
    <au><snm>Bakerly</snm><fnm>ND</fnm></au>
    <au><snm>Singh</snm><fnm>D</fnm></au>
    <au><snm>Williamson</snm><fnm>PR</fnm></au>
    <au><snm>Felton</snm><fnm>T</fnm></au>
    <au><cnm>others</cnm></au>
  </aug>
  <source>Life</source>
  <publisher>MDPI</publisher>
  <pubdate>2020</pubdate>
  <volume>10</volume>
  <issue>12</issue>
  <fpage>350</fpage>
</bibl>

<bibl id="B4">
  <title><p>Estimation of the probability of an event as a function of several independent variables</p></title>
  <aug>
    <au><snm>Walker</snm><fnm>SH</fnm></au>
    <au><snm>Duncan</snm><fnm>DB</fnm></au>
  </aug>
  <source>Biometrika</source>
  <publisher>Oxford University Press</publisher>
  <pubdate>1967</pubdate>
  <volume>54</volume>
  <issue>1-2</issue>
  <fpage>167</fpage>
  <lpage>-179</lpage>
</bibl>

<bibl id="B5">
  <title><p>Regression models for ordinal data</p></title>
  <aug>
    <au><snm>McCullagh</snm><fnm>P</fnm></au>
  </aug>
  <source>Journal of the Royal Statistical Society: Series B (Methodological)</source>
  <publisher>Wiley Online Library</publisher>
  <pubdate>1980</pubdate>
  <volume>42</volume>
  <issue>2</issue>
  <fpage>109</fpage>
  <lpage>-127</lpage>
</bibl>

<bibl id="B6">
  <title><p>Regression modeling strategies: with applications to linear models, logistic and ordinal regression, and survival analysis</p></title>
  <aug>
    <au><snm>Harrell</snm><fnm>FE</fnm></au>
    <au><cnm>others</cnm></au>
  </aug>
  <publisher>Springer</publisher>
  <pubdate>2015</pubdate>
  <volume>3</volume>
</bibl>

<bibl id="B7">
  <title><p>Efficient adaptive designs for clinical trials of interventions for COVID-19</p></title>
  <aug>
    <au><snm>Stallard</snm><fnm>N</fnm></au>
    <au><snm>Hampson</snm><fnm>L</fnm></au>
    <au><snm>Benda</snm><fnm>N</fnm></au>
    <au><snm>Brannath</snm><fnm>W</fnm></au>
    <au><snm>Burnett</snm><fnm>T</fnm></au>
    <au><snm>Friede</snm><fnm>T</fnm></au>
    <au><snm>Kimani</snm><fnm>PK</fnm></au>
    <au><snm>Koenig</snm><fnm>F</fnm></au>
    <au><snm>Krisam</snm><fnm>J</fnm></au>
    <au><snm>Mozgunov</snm><fnm>P</fnm></au>
    <au><cnm>others</cnm></au>
  </aug>
  <source>Statistics in Biopharmaceutical Research</source>
  <publisher>Taylor \& Francis</publisher>
  <pubdate>2020</pubdate>
  <volume>12</volume>
  <issue>4</issue>
  <fpage>483</fpage>
  <lpage>-497</lpage>
</bibl>

<bibl id="B8">
  <title><p>RECOVERY-respiratory support: respiratory strategies for patients with suspected or proven COVID-19 respiratory failure; continuous positive airway pressure, high-flow nasal oxygen, and standard care: a structured summary of a study protocol for a randomised controlled trial</p></title>
  <aug>
    <au><snm>Perkins</snm><fnm>GD</fnm></au>
    <au><snm>Couper</snm><fnm>K</fnm></au>
    <au><snm>Connolly</snm><fnm>B</fnm></au>
    <au><snm>Baillie</snm><fnm>JK</fnm></au>
    <au><snm>Bradley</snm><fnm>JM</fnm></au>
    <au><snm>Dark</snm><fnm>P</fnm></au>
    <au><snm>De Soyza</snm><fnm>A</fnm></au>
    <au><snm>Gorman</snm><fnm>E</fnm></au>
    <au><snm>Gray</snm><fnm>A</fnm></au>
    <au><snm>Hamilton</snm><fnm>L</fnm></au>
    <au><cnm>others</cnm></au>
  </aug>
  <source>Trials</source>
  <publisher>Springer</publisher>
  <pubdate>2020</pubdate>
  <volume>21</volume>
  <issue>1</issue>
  <fpage>1</fpage>
  <lpage>-3</lpage>
</bibl>

<bibl id="B9">
  <title><p>Effect of antiplatelet therapy on survival and organ support--free days in critically ill patients with COVID-19: a randomized clinical trial</p></title>
  <aug>
    <au><snm>Florescu</snm><fnm>S</fnm></au>
    <au><snm>Stanciu</snm><fnm>D</fnm></au>
    <au><snm>Zaharia</snm><fnm>M</fnm></au>
    <au><snm>Kosa</snm><fnm>A</fnm></au>
    <au><snm>Codreanu</snm><fnm>D</fnm></au>
    <au><snm>Kidwai</snm><fnm>A</fnm></au>
    <au><snm>Masood</snm><fnm>S</fnm></au>
    <au><snm>Kaye</snm><fnm>C</fnm></au>
    <au><snm>Coutts</snm><fnm>A</fnm></au>
    <au><snm>MacKay</snm><fnm>L</fnm></au>
    <au><cnm>others</cnm></au>
  </aug>
  <source>Jama</source>
  <publisher>American Medical Association</publisher>
  <pubdate>2022</pubdate>
  <volume>327</volume>
  <issue>13</issue>
  <fpage>1247</fpage>
  <lpage>-1259</lpage>
</bibl>

<bibl id="B10">
  <title><p>Adaptive trial designs: a review of barriers and opportunities</p></title>
  <aug>
    <au><snm>Kairalla</snm><fnm>JA</fnm></au>
    <au><snm>Coffey</snm><fnm>CS</fnm></au>
    <au><snm>Thomann</snm><fnm>MA</fnm></au>
    <au><snm>Muller</snm><fnm>KE</fnm></au>
  </aug>
  <source>Trials</source>
  <publisher>Springer</publisher>
  <pubdate>2012</pubdate>
  <volume>13</volume>
  <fpage>1</fpage>
  <lpage>-9</lpage>
</bibl>

<bibl id="B11">
  <title><p>Angiotensin--neprilysin inhibition versus enalapril in heart failure</p></title>
  <aug>
    <au><snm>McMurray</snm><fnm>JJ</fnm></au>
    <au><snm>Packer</snm><fnm>M</fnm></au>
    <au><snm>Desai</snm><fnm>AS</fnm></au>
    <au><snm>Gong</snm><fnm>J</fnm></au>
    <au><snm>Lefkowitz</snm><fnm>MP</fnm></au>
    <au><snm>Rizkala</snm><fnm>AR</fnm></au>
    <au><snm>Rouleau</snm><fnm>JL</fnm></au>
    <au><snm>Shi</snm><fnm>VC</fnm></au>
    <au><snm>Solomon</snm><fnm>SD</fnm></au>
    <au><snm>Swedberg</snm><fnm>K</fnm></au>
    <au><cnm>others</cnm></au>
  </aug>
  <source>N Engl J Med</source>
  <publisher>Mass Medical Soc</publisher>
  <pubdate>2014</pubdate>
  <volume>371</volume>
  <fpage>993</fpage>
  <lpage>-1004</lpage>
</bibl>

<bibl id="B12">
  <title><p>Bayesian clinical trials</p></title>
  <aug>
    <au><snm>Berry</snm><fnm>DA</fnm></au>
  </aug>
  <source>Nature reviews Drug discovery</source>
  <publisher>Nature Publishing Group UK London</publisher>
  <pubdate>2006</pubdate>
  <volume>5</volume>
  <issue>1</issue>
  <fpage>27</fpage>
  <lpage>-36</lpage>
</bibl>

<bibl id="B13">
  <title><p>Holland-Frei cancer medicine 8</p></title>
  <aug>
    <au><snm>Bast Jr</snm><fnm>RC</fnm></au>
    <au><snm>Holland</snm><fnm>JF</fnm></au>
  </aug>
  <publisher>PMPH-USA</publisher>
  <pubdate>2010</pubdate>
</bibl>

<bibl id="B14">
  <title><p>Introduction to Bayesian methods III: use and interpretation of Bayesian tools in design and analysis</p></title>
  <aug>
    <au><snm>Berry</snm><fnm>DA</fnm></au>
  </aug>
  <source>Clinical Trials</source>
  <publisher>Sage Publications Sage CA: Thousand Oaks, CA</publisher>
  <pubdate>2005</pubdate>
  <volume>2</volume>
  <issue>4</issue>
  <fpage>295</fpage>
  <lpage>-300</lpage>
</bibl>

<bibl id="B15">
  <title><p>Philosophy and the practice of Bayesian statistics</p></title>
  <aug>
    <au><snm>Gelman</snm><fnm>A</fnm></au>
    <au><snm>Shalizi</snm><fnm>CR</fnm></au>
  </aug>
  <source>British Journal of Mathematical and Statistical Psychology</source>
  <publisher>Wiley Online Library</publisher>
  <pubdate>2013</pubdate>
  <volume>66</volume>
  <issue>1</issue>
  <fpage>8</fpage>
  <lpage>-38</lpage>
</bibl>

<bibl id="B16">
  <title><p>Bayesian model choice in cumulative link ordinal regression models</p></title>
  <aug>
    <au><snm>McKinley</snm><fnm>TJ</fnm></au>
    <au><snm>Morters</snm><fnm>M</fnm></au>
    <au><snm>Wood</snm><fnm>JL</fnm></au>
  </aug>
  <pubdate>2015</pubdate>
</bibl>

<bibl id="B17">
  <title><p>Bayesian Cumulative Probability Models for Continuous and Mixed Outcomes</p></title>
  <aug>
    <au><snm>James</snm><fnm>NT</fnm></au>
    <au><snm>Harrell Jr</snm><fnm>FE</fnm></au>
    <au><snm>Shepherd</snm><fnm>BE</fnm></au>
  </aug>
  <source>arXiv preprint arXiv:2102.00330</source>
  <pubdate>2021</pubdate>
</bibl>

<bibl id="B18">
  <title><p>Anticoagulation strategies in non--critically ill patients with Covid-19</p></title>
  <aug>
    <au><snm>McQuilten</snm><fnm>ZK</fnm></au>
    <au><snm>Venkatesh</snm><fnm>B</fnm></au>
    <au><snm>Jha</snm><fnm>V</fnm></au>
    <au><snm>Roberts</snm><fnm>J</fnm></au>
    <au><snm>Morpeth</snm><fnm>SC</fnm></au>
    <au><snm>Totterdell</snm><fnm>JA</fnm></au>
    <au><snm>McPhee</snm><fnm>GM</fnm></au>
    <au><snm>Abraham</snm><fnm>J</fnm></au>
    <au><snm>Bam</snm><fnm>N</fnm></au>
    <au><snm>Bandara</snm><fnm>M</fnm></au>
    <au><cnm>others</cnm></au>
  </aug>
  <source>NEJM Evidence</source>
  <publisher>Massachusetts Medical Society</publisher>
  <pubdate>2023</pubdate>
  <volume>2</volume>
  <issue>2</issue>
  <fpage>EVIDoa2200293</fpage>
</bibl>

<bibl id="B19">
  <title><p>Bayesian data analysis</p></title>
  <aug>
    <au><snm>Gelman</snm><fnm>A</fnm></au>
    <au><snm>Carlin</snm><fnm>JB</fnm></au>
    <au><snm>Stern</snm><fnm>HS</fnm></au>
    <au><snm>Rubin</snm><fnm>DB</fnm></au>
  </aug>
  <publisher>Chapman and Hall/CRC</publisher>
  <pubdate>2014</pubdate>
</bibl>

<bibl id="B20">
  <title><p>A weakly informative default prior distribution for logistic and other regression models</p></title>
  <aug>
    <au><snm>Gelman</snm><fnm>A</fnm></au>
    <au><snm>Jakulin</snm><fnm>A</fnm></au>
    <au><snm>Pittau</snm><fnm>MG</fnm></au>
    <au><snm>Su</snm><fnm>YS</fnm></au>
  </aug>
  <pubdate>2008</pubdate>
</bibl>

<bibl id="B21">
  <title><p>Prior setting in practice: Strategies and rationales used in choosing prior distributions for Bayesian analysis</p></title>
  <aug>
    <au><snm>Sarma</snm><fnm>A</fnm></au>
    <au><snm>Kay</snm><fnm>M</fnm></au>
  </aug>
  <source>Proceedings of the 2020 chi conference on human factors in computing systems</source>
  <pubdate>2020</pubdate>
  <fpage>1</fpage>
  <lpage>-12</lpage>
</bibl>

<bibl id="B22">
  <title><p>Prior Distributions for Bayesian Regression Using PyMC</p></title>
  <aug>
    <au><snm>Rochford</snm><fnm>A</fnm></au>
  </aug>
  <source>\url{https://austinrochford.com/posts/2013-09-02-prior-distributions-for-bayesian-regression-using-pymc.html}</source>
  <pubdate>2013</pubdate>
  <note>Accessed: 2025-01-20</note>
</bibl>

<bibl id="B23">
  <title><p>Estimating Regularized Linear Models with rstanarm</p></title>
  <aug>
    <au><snm>Gabry</snm><fnm>J</fnm></au>
    <au><snm>Goodrich</snm><fnm>B</fnm></au>
  </aug>
  <source>\url{https://cran.r-project.org/web/packages/rstanarm/vignettes/lm.html#priors}</source>
  <pubdate>2024</pubdate>
  <note>Accessed: 2025-01-20</note>
</bibl>

<bibl id="B24">
  <title><p>The dirichlet-multinomial and dirichlet-categorical models for bayesian inference</p></title>
  <aug>
    <au><snm>Tu</snm><fnm>S</fnm></au>
  </aug>
  <source>Computer Science Division, UC Berkeley</source>
  <pubdate>2014</pubdate>
  <volume>2</volume>
</bibl>

<bibl id="B25">
  <title><p>Efficacy and safety of nerinetide for the treatment of acute ischaemic stroke (ESCAPE-NA1): a multicentre, double-blind, randomised controlled trial</p></title>
  <aug>
    <au><snm>Hill</snm><fnm>MD</fnm></au>
    <au><snm>Goyal</snm><fnm>M</fnm></au>
    <au><snm>Menon</snm><fnm>BK</fnm></au>
    <au><snm>Nogueira</snm><fnm>RG</fnm></au>
    <au><snm>McTaggart</snm><fnm>RA</fnm></au>
    <au><snm>Demchuk</snm><fnm>AM</fnm></au>
    <au><snm>Poppe</snm><fnm>AY</fnm></au>
    <au><snm>Buck</snm><fnm>BH</fnm></au>
    <au><snm>Field</snm><fnm>TS</fnm></au>
    <au><snm>Dowlatshahi</snm><fnm>D</fnm></au>
    <au><cnm>others</cnm></au>
  </aug>
  <source>The Lancet</source>
  <publisher>Elsevier</publisher>
  <pubdate>2020</pubdate>
  <volume>395</volume>
  <issue>10227</issue>
  <fpage>878</fpage>
  <lpage>-887</lpage>
</bibl>

<bibl id="B26">
  <title><p>R: A Language and Environment for Statistical Computing</p></title>
  <aug>
    <au><cnm>{R Core Team}</cnm></au>
  </aug>
  <publisher>Vienna, Austria</publisher>
  <pubdate>2021</pubdate>
  <url>https://www.R-project.org/</url>
</bibl>

<bibl id="B27">
  <title><p>Stan: A probabilistic programming language</p></title>
  <aug>
    <au><snm>Carpenter</snm><fnm>B</fnm></au>
    <au><snm>Gelman</snm><fnm>A</fnm></au>
    <au><snm>Hoffman</snm><fnm>MD</fnm></au>
    <au><snm>Lee</snm><fnm>D</fnm></au>
    <au><snm>Goodrich</snm><fnm>B</fnm></au>
    <au><snm>Betancourt</snm><fnm>M</fnm></au>
    <au><snm>Brubaker</snm><fnm>M</fnm></au>
    <au><snm>Guo</snm><fnm>J</fnm></au>
    <au><snm>Li</snm><fnm>P</fnm></au>
    <au><snm>Riddell</snm><fnm>A</fnm></au>
  </aug>
  <source>Journal of statistical software</source>
  <publisher>Columbia Univ., New York, NY (United States); Harvard Univ., Cambridge, MA~…</publisher>
  <pubdate>2017</pubdate>
  <volume>76</volume>
  <issue>1</issue>
</bibl>

<bibl id="B28">
  <title><p>posterior: Tools for Working with Posterior Distributions</p></title>
  <aug>
    <au><snm>Bürkner</snm><fnm>PC</fnm></au>
    <au><snm>Gabry</snm><fnm>J</fnm></au>
    <au><snm>Kay</snm><fnm>M</fnm></au>
    <au><snm>Vehtari</snm><fnm>A</fnm></au>
  </aug>
  <pubdate>2024</pubdate>
  <url>https://mc-stan.org/posterior/</url>
  <note>R package version 1.6.0</note>
</bibl>

<bibl id="B29">
  <title><p>rstanarm: {Bayesian} applied regression modeling via {Stan}</p></title>
  <aug>
    <au><snm>Goodrich</snm><fnm>B</fnm></au>
    <au><snm>Gabry</snm><fnm>J</fnm></au>
    <au><snm>Ali</snm><fnm>I</fnm></au>
    <au><snm>Brilleman</snm><fnm>S</fnm></au>
  </aug>
  <pubdate>2020</pubdate>
  <url>https://mc-stan.org/rstanarm</url>
</bibl>

<bibl id="B30">
  <title><p>The Bayesian simulation study (BASIS) framework for simulation studies in statistical and methodological research</p></title>
  <aug>
    <au><snm>Kelter</snm><fnm>R</fnm></au>
  </aug>
  <source>Biometrical Journal</source>
  <publisher>Wiley Online Library</publisher>
  <pubdate>2023</pubdate>
  <fpage>2200095</fpage>
</bibl>

<bibl id="B31">
  <title><p>On the assessment of Monte Carlo error in simulation-based statistical analyses</p></title>
  <aug>
    <au><snm>Koehler</snm><fnm>E</fnm></au>
    <au><snm>Brown</snm><fnm>E</fnm></au>
    <au><snm>Haneuse</snm><fnm>SJP</fnm></au>
  </aug>
  <source>The American Statistician</source>
  <publisher>Taylor \& Francis</publisher>
  <pubdate>2009</pubdate>
  <volume>63</volume>
  <issue>2</issue>
  <fpage>155</fpage>
  <lpage>-162</lpage>
</bibl>

<bibl id="B32">
  <title><p>Using simulation studies to evaluate statistical methods</p></title>
  <aug>
    <au><snm>Morris</snm><fnm>TP</fnm></au>
    <au><snm>White</snm><fnm>IR</fnm></au>
    <au><snm>Crowther</snm><fnm>MJ</fnm></au>
  </aug>
  <source>Statistics in medicine</source>
  <publisher>Wiley Online Library</publisher>
  <pubdate>2019</pubdate>
  <volume>38</volume>
  <issue>11</issue>
  <fpage>2074</fpage>
  <lpage>-2102</lpage>
</bibl>

<bibl id="B33">
  <title><p>rstanarm: Bayesian applied regression modeling via Stan</p></title>
  <aug>
    <au><snm>Goodrich</snm><fnm>B</fnm></au>
    <au><snm>Gabry</snm><fnm>J</fnm></au>
    <au><snm>Ali</snm><fnm>I</fnm></au>
    <au><snm>Brilleman</snm><fnm>S</fnm></au>
  </aug>
  <source>R package version</source>
  <pubdate>2020</pubdate>
  <volume>2</volume>
  <issue>1</issue>
</bibl>

<bibl id="B34">
  <title><p>Rank-normalization, folding, and localization: An improved R ̂ for assessing convergence of MCMC (with discussion)</p></title>
  <aug>
    <au><snm>Vehtari</snm><fnm>A</fnm></au>
    <au><snm>Gelman</snm><fnm>A</fnm></au>
    <au><snm>Simpson</snm><fnm>D</fnm></au>
    <au><snm>Carpenter</snm><fnm>B</fnm></au>
    <au><snm>B{\"u}rkner</snm><fnm>PC</fnm></au>
  </aug>
  <source>Bayesian analysis</source>
  <publisher>International Society for Bayesian Analysis</publisher>
  <pubdate>2021</pubdate>
  <volume>16</volume>
  <issue>2</issue>
  <fpage>667</fpage>
  <lpage>-718</lpage>
</bibl>

<bibl id="B35">
  <title><p>Randomised trials with provision for early stopping for benefit (or harm): the impact on the estimated treatment effect</p></title>
  <aug>
    <au><snm>Walter</snm><fnm>SD</fnm></au>
    <au><snm>Guyatt</snm><fnm>GH</fnm></au>
    <au><snm>Bassler</snm><fnm>D</fnm></au>
    <au><snm>Briel</snm><fnm>M</fnm></au>
    <au><snm>Ramsay</snm><fnm>T</fnm></au>
    <au><snm>Han</snm><fnm>HD</fnm></au>
  </aug>
  <source>Statistics in medicine</source>
  <publisher>Wiley Online Library</publisher>
  <pubdate>2019</pubdate>
  <volume>38</volume>
  <issue>14</issue>
  <fpage>2524</fpage>
  <lpage>-2543</lpage>
</bibl>

<bibl id="B36">
  <title><p>Centring in regression analyses: a strategy to prevent errors in statistical inference</p></title>
  <aug>
    <au><snm>Kraemer</snm><fnm>HC</fnm></au>
    <au><snm>Blasey</snm><fnm>CM</fnm></au>
  </aug>
  <source>International journal of methods in psychiatric research</source>
  <publisher>Wiley Online Library</publisher>
  <pubdate>2004</pubdate>
  <volume>13</volume>
  <issue>3</issue>
  <fpage>141</fpage>
  <lpage>-151</lpage>
</bibl>

<bibl id="B37">
  <title><p>Simulation Study Code</p></title>
  <aug>
    <au><snm>Selman</snm><fnm>CJ</fnm></au>
  </aug>
  <source>\url{https://github.com/chrisselman/noninfo_pomodel}</source>
  <publisher>GitHub</publisher>
  <pubdate>2025</pubdate>
</bibl>

</refgrp>
} 


\end{backmatter}

\end{document}